\documentclass[onecolumn,superscriptaddress,floatfix,preprintnumbers,nofootinbib]{revtex4-2}
\usepackage{lipsum}
\usepackage[english]{babel}

\usepackage{amssymb}
\usepackage{amsmath}
\usepackage{graphicx}
\usepackage{graphics}
\usepackage{xcolor}
\usepackage{booktabs} % LT for tables
\usepackage{physics} % LT for useful routines
\usepackage{siunitx} % For SI stuff
\usepackage{empheq}
\DeclareSIUnit \parsec {pc}
\DeclareSIUnit \year {yr}

\usepackage{hyperref}

\newcommand{\e}{\mathrm{e}}
\begin{document}
\title{The approximate gravitational lensing multiple plane mass sheet degeneracy}
 
 \author{Luca Teodori}
 \email{luca.teodori@iac.es}  
 \affiliation{Instituto de Astrof\'isica de Canarias, C/ V\'ia L\'actea, s/n E38205, La Laguna, Tenerife, Spain
 } 
 \affiliation{Universidad de La Laguna, Departamento de Astrof\'isica, La Laguna, Tenerife,
 	Spain}
\begin{abstract}
Strong gravitational lensing has to deal with many modeling degeneracies, the most notable being the Mass Sheet Degeneracy (MSD). We review the MSD when one needs to model more lens planes, each one with an internal mass sheet. We take into account the non-linear lens-lens coupling and line of sight effects, the latter treated as external mass sheets with associated shear. If second order shear terms on external and internal mass sheets can be neglected, we show that the MSD is always retained, and the mass sheets influence can be reabsorbed in the redefinition of angular diameter distances. In particular, internal and external mass sheets can be placed on the same footing. The version of the MSD discussed here does not require any particular relation between the internal mass sheets in the different planes. Even when including time delays from all sources, a residual degeneracy involving time delays, mass sheets and $ H_0 $ remains. We develop a framework which shows what can actually be constrained in multiple plane lens systems. %Since effective angular diameter distances contain both cosmology and mass sheets, %In particular, the apparent breaking of MSD from multiple source systems may be absorbed on the unobservable external convergences. 
%using angular diameter distance ratio to constrain cosmological parameters while not considering the generic version of MSD we here discuss can bias their inference.
%
\end{abstract}
%\pacs{}
\maketitle
%\tableofcontents
%%%%%%%%%%%%%%%%%%%
\section{Introduction} \label{s:intro}
Gravitational lensing offers an unique probe in the structure of galaxies~\cite{Vegetti:2023mgp}, the distribution of dark matter, and cosmology~\cite{Refsdal:1964nw,Millon:2019slk,Grillo:2024rhi}. Features of lensed images allow modelers to reconstruct the lens density (or, more correctly, the column density) model. Moreover, gravitational lensing can be used as a tool to measure cosmological parameters, like the matter density parameter $ \Omega_{\rm m} $ or the Hubble constant $ H_0 $~\cite{Birrer:2018vtm,Birrer:2020tax,TDCOSMO:2025dmr,Kelly:2023mgv,Grillo:2024rhi,Collett:2014ola,Sharma:2022xyw,Bowden:2025rph}.

The reconstruction of the lens density model from imaging data suffers a well-known degeneracy, called the Mass Sheet Degeneracy (MSD)~\cite{Falco1985,Schneider:2013sxa}, which is a special case of the more general source-position transformation~\cite{Schneider:2014vka}\footnote{A generic source position transformation degeneracy is broken by time delays, independently of the knowledge of $ H_0 $, unless the source in the transformed model is proportional to the source of the original model via a constant, i.e. if it is a MSD. Since our primary motivation is the investigation of degeneracies in the measurement of $ H_0 $ from strong gravitational lensing systems, we limit the discussion here to the MSD alone.}. 
Physical realizations of column densities that cause the MSD may arise from matter on the Line of Sight (LOS)~\cite{Suyu:2009by,Fleury:2021tke}, from unaccounted cores~\cite{Schneider:2013sxa,Blum:2020mgu,Blum:2021oxj}, or from the presence of a nearby group~\cite{Wilson:2016hcs,Wilson:2017apg,Teodori:2023nrz}. Matter along the LOS that can cause a MSD is usually referred to as external mass sheets, as its mass is distributed at redshifts different from that of the main lens one, whereas a MSD coming from column densities located at the lens redshift can be said to originate from internal mass sheets. This distinction is important, as they affect observables such as stellar kinematics in a different way. But at the same time, both types of mass sheets are fundamentally connected. We emphasize that, to form a MSD, such matter distributions need to resemble a ``sheet'' only in projection.

%Examples of observables which can break the MSD are absolute magnifications (which however are hard to come by, even when using standard candles like Supernovae Ia) and time delays. 
If cosmology is assumed, in particular if the Hubble constant $ H_0 $ is considered known, then time delays between multiple images of a gravitational lens can break the MSD. However, if time delays are used to measure the Hubble constant, then the degeneracy persists, and ignoring the MSD leads to systematic biases in such a measurement. 
Given the current tension in measurements of $ H_0 $~\cite{DiValentino:2021izs,Verde:2023lmm,DiValentino:2025sru} between early universe probes like the Cosmic Microwave Background (CMB) and Large Scale Structures (LSS)~\cite{Akrami:2018vks,Abuter:2018drb,Ivanov:2019pdj,Troster:2019ean} and distance-ladder based methods (see Ref.~\cite{Riess:2021jrx} for the Cepheid-calibrated type Ia supernovae measurement, although, for tip of the red giant branch-calibrated supernovae Ia, the tension is milder~\cite{Freedman:2024eph}), there is interest in gravitational lensing time delays, as they offer an alternative, independent way to measure $ H_0 $. A careful treatment of degeneracies that can affect the accuracy of such a measurement, and how they can be broken or mitigated, is therefore important.

The MSD can be broken if absolute lensing magnifications are available~\cite{Kolatt:1997zh,Oguri:2002ku}, or if the mass of the lens is known~\cite{Grogin:1995gf,Romanowsky_1999,TreuKoopmans02}. However, in most cases (such as quasar lensing), the absolute magnification is unknown, and even when absolute magnifications can in principle be accessed (as in the case of supernovae Ia), image fluxes are usually contaminated by stellar microlensing (see e.g. Ref.~\cite{Foxley-Marrable:2018dzu}). As a result, fluxes are not used in most lensing pipelines. To instead obtain a prior on the mass of the lens, a widely used observable comes from stellar kinematics measurements within the lens galaxy.
As far as stellar kinematics is concerned, Ref.~\cite{Birrer:2020jyr} forecasts a percent-level accurate inference of $ H_0 $ using resolved stellar kinematics of many gravitational lens systems. As of now, efforts to measure $ H_0 $ using massive ellipticals come primarily from the TDCOSMO collaboration~\cite{Millon:2019slk}, where stellar kinematics plays an important role in mitigating the MSD. However, the way stellar kinematics breaks the degeneracy is not immune to additional systematics, like the mass-anisotropy degeneracy~\cite{1987ApJ...313..121M,Mamon:2005}. It is therefore important to explore as many ways as possible for breaking the MSD. 

The use of multiple source systems, or multiple plane systems, might at a first sight be considered as a possible other way to break the MSD. Although only a relatively small number of non-cluster multiple plane systems are currently known, with LSST~\cite{LSSTScience:2009jmu,LSSTDarkEnergyScience:2018jkl} and Euclid~\cite{Euclid:2019clj} we expect $ \mathcal{O}(1000) $ new multiple plane systems to be discovered~\cite{Collett:2015roa,Sharma:2022xyw}, making their study timely. It is important to note that versions of the MSD also exist in the multiple plane case, as shown in~\cite{Schneider:2014ifa,Schneider:2014vka}. However, such a form of the MSD requires alternating mass sheets on different planes, tuned as to produce an exact degeneracy. The need for such fine-tuning is sometimes used as an argument that multiple plane systems effectively break the MSD (see e.g. claim in Refs.~\cite{Schneider:2014vka,Dux:2024vvq}). However, in the MSD presented in Ref.~\cite{Schneider:2014vka}, the fine-tuning between different mass sheets is needed because it is required for the mathematical degeneracy to be exact, \emph{and it is assumed that angular diameter distances are untouched by the degeneracy}.

Interestingly, angular diameter distance ratios are observable in multiple plane systems, and this possibility has been used to constrain cosmology~\cite{Collett:2014ola,Ballard:2023fgi,Bowden:2025rph}.
However, the actual observable in multiple plane systems are, as we will show, what we refer to here as effective angular diameter distances, which are composed of the Friedmann-Robertson-Walker-Lemaitre (FRWL) angular diameter distances and corrections arising from unaccounted mass sheets, \textit{both internal and external}. This point is crucial, as it implies that angular diameter distances must enter the MSD. Assuming that the FRWL angular diameter distances are directly observable can lead to biases in the inference of cosmological parameters.
 
In Ref.~\cite{Teodori:2022ltt}, we investigated the angular diameter distance ratio, focusing on the case in which lens-lens coupling is neglected. We showed that the apparent breaking of the MSD due to an internal mass sheet in one lens plane can be masked by differences in external convergences between the different planes. We argued that, if the internal mass sheet effect is larger than what is expected from external cosmological convergences, then multiple sources can help in spotting the internal mass sheet.

In this work, we re-examine the multiple plane MSD, including the lens-lens coupling and LOS effect with shear, with a focus on the effective angular diameter distance ratios. We describe our framework for dealing with a generalized multiple plane MSD and clarify what effective angular diameter distance ratios actually constrain. In particular, the use of effective angular diameter distances to express the lens and time delay equations makes the passage between models connected by a MSD almost trivial.
We show that the multiple plane MSD, in which no constraints are imposed on the mass sheets internal to the various lens planes (unlike in Ref.~\cite{Schneider:2014vka}), is still retained when convergences of internal mass sheets have values comparable to external ones. In fact, there is no real mathematical difference between internal and external sheets. Even with (unlikely) knowledge of the time delays from all sources within a multiple plane lens system, the degeneracy between mass sheets and $ H_0 $ persists. We conclude that no fine-tuning is needed for the multiple plane MSD to be retained, and we thus emphasize once again that multiple plane systems do not really break the MSD.
In Table~\ref{tab:summary}, we summarize which quantities are observable and the types of degeneracies they are involved in within multiple source systems, in light of the MSD discussed in this work.
%The version of the multiple plane MSD we will present here is more general than all the others found in the literature CITE.

This work is structured as follows. We describe the multiple plane lens equation, together with our notation, in Sec.~\ref{s:multi_plane} and its MSD-reduced version in Sec.~\ref{s:msd}. We discuss how time delays change within the original and MSD-reduced model in Sec.~\ref{s:time_delay}. We give an example on how this MSD can affect galaxy multiple plane systems in Sec.~\ref{s:cluster}; we explore how higher order effects affect our discussion in Sec.~\ref{s:non_linear}; we compare our framework with previous works in Sec.~\ref{s:discussion}; we conclude in Sec.~\ref{s:sum}.

\section{The multiple plane lens equation with external and internal sheets} \label{s:multi_plane}
Consider a gravitational lens system with $ N+1 $ planes, with $ n=0,\ldots,N $ denoting the $ n $-th plane, where $ n=0 $ denotes the first plane, or (main) lens plane. The $ N $-th plane will be the furthest source plane. Inserting LOS effects within the tidal approximation, the recursive multiple plane lens equation reads~\cite{Schneider:1992,McCully:2013fga,Schneider:2014vka,Teodori:2022ltt} (see also App.~\ref{s:der})
\begin{equation} \label{eq:multiplane}
%	\vec\beta_i = (1-\kappa(z_i,0))(\mathbb{I} + \Gamma(z_i,0)) \vec\theta - \sum_{j=0}^{i-1} (1-\kappa(z_i,z_j))(\mathbb{I} + \Gamma(z_i,z_j)) C_{ij} (\vec\alpha_j(\vec\beta_j) +M_{\mathrm{c}j}\vec\beta_j) \ , 
\vec\beta_n = \qty[\mathbb{I}-\kappa(z_n,0) - \Gamma(z_n,0)]\vec\theta - \sum_{m=0}^{n-1} \qty[\mathbb{I}-\kappa(z_n,z_m) - \Gamma(z_n,z_m)] C_{nm} \qty(\vec\alpha_m(\vec\beta_m) +(\kappa_{\mathrm{c}m} + \Gamma_{\mathrm{c}m})\vec\beta_m) \ , 
\end{equation}
where
\begin{equation}
C_{nm} := \frac{D_{\rm A}(z_{N},0)  D_{\rm A} (z_n, z_m)}{ D_{\rm A}(z_{n},0)  D_{\rm A}(z_{N}, z_m)} \ ,
\end{equation}
is the angular diameter distance ratio, coming from the following choice to relate the $ n $-th plane deflection angle $\hat{\alpha}_n$ with the $ n $-th plane displacement angle $ \vec{\alpha}_n $~\cite{Schneider:1992}
\begin{equation} \label{eq:choice}
	\vec{\alpha}_n = \frac{D_{\rm A}(z_{N},z_n)}{D_{\rm A}(z_{N},0)} \hat{\alpha}_n \ .
\end{equation}
$ \vec{\theta} $ is the observed angle, while $ \vec{\beta}_n $ is the angle on the $ n $-th plane the light ray forms when crossing the $ n $-th plane. $ \vec{\beta}_N $ is the angle of the final source with respect to its plane.

$ D_{\rm A}(z_n,z_m) $, $ \kappa(z_n, z_m) $ and $ \Gamma(z_n,z_m) $ are the FRWL angular diameter distance, external convergence and the shear matrix respectively, from redshift $ z_m $ to redshift $ z_n $, with
\begin{equation}
%M(z_i,z_j) := \begin{pmatrix}
%	\kappa(z_i, z_j) + \gamma_1(z_i, z_j) & \gamma_2(z_i, z_j) \\
%	\gamma_2(z_i, z_j) & \kappa(z_i, z_j) - \gamma_1(z_i, z_j) 
%\end{pmatrix} \ .
\Gamma(z_n,z_m) = \begin{pmatrix}
	 \gamma_1(z_n, z_m) & \gamma_2(z_n, z_m) \\
	\gamma_2(z_n, z_m) & - \gamma_1(z_n, z_m) 
\end{pmatrix} \ .
\end{equation}
$ \kappa_{\mathrm{c} m} $ and $ \Gamma_{\mathrm{c}m} $ represent possible internal mass sheets (with an associated constant convergence $ \kappa_{\mathrm{c} m} $ and shear $ \Gamma_{\mathrm{c} m} $) on lens plane $ m $. {\bm We use the subscript $ _{\rm c} $ to denote these quantities, as in physical realizations of internal MSD they are connected to cores or profiles close to cores}. Thus, Eq.~\eqref{eq:multiplane} explicitly contains external (to all the lens planes) mass sheets from cosmological contributions\footnote{Such cosmological contributions can in principle be estimated by ray-tracing methods on cosmological simulations, on the lines of what shown in Refs.~\cite{Hilbert:2008kb,Johnson:2025lhy}.} and internal mass sheets, coming from column density contributions lying at the same redshift of the various lens planes (or, in other words, coming from the lens models).

\section{The multiple plane MSD} \label{s:msd}
We want to rewrite Eq.~\eqref{eq:multiplane} by reabsorbing all the mass sheet terms $ \kappa(z_n, z_m) $ and $ \kappa_{\mathrm{c}m} $, in the form
\begin{equation} \label{eq:msd_multi}
\vec{\tilde{\beta}}_n = (\mathbb{I} - \tilde\Gamma_n)\vec{\theta} - \sum_{m=0}^{n-1} (\mathbb{I} - \tilde{\Gamma}_{nm}) \tilde{C}_{nm} \vec{\tilde{\alpha}}_m(\vec{\tilde{\beta}}_m) \ , 
\end{equation}
where in the following we employ the abbreviations $ \kappa_{nm} := \kappa(z_n,z_m) $, $ \kappa_{n} := \kappa(z_n,0) $ and  $ \Gamma_{nm} := \Gamma(z_n,z_m) $, $ \Gamma_{n} := \Gamma(z_n,0) $. We call the model in Eq.~\eqref{eq:msd_multi} the MSD-reduced model.
We require the quantities in Eq.~\eqref{eq:msd_multi} to be connected to the quantities in Eq.~\eqref{eq:multiplane} with constants $ \lambda $s and, for the shear matrices, shifts $ \Gamma^{(\mathrm{c})}_{nm}  $, $ \Gamma^{(\mathrm{c})}_n $ (subscript $ ^{(\mathrm{c})} $ stands again for core, as these shifts will be related to $ \Gamma_{\mathrm{c}n} $ terms).
\begin{equation} \label{eq:lambdas}
\vec{\tilde{\beta}}_n=\lambda_n \vec\beta_n \ , \ \tilde{C}_{nm} = \lambda_{nm} C_{nm} \ , \ \tilde{\alpha}_n(\vec\beta) = \lambda^\alpha_{n} \vec\alpha(\lambda^{\rm L}_n\vec\beta) \ , \ \tilde\Gamma(z_n,z_m) = \lambda^{\Gamma}_{nm} \Gamma_k(z_n,z_m) + \Gamma^{(\mathrm{c})}_{nm} \ , \ \tilde\Gamma(z_n,0) = \lambda^{\Gamma}_{n} \Gamma_k(z_n,0) + \Gamma^{(\mathrm{c})}_{n}  \ .
\end{equation}
It turns out that, for generic $ \Gamma_{\mathrm{c}m} $, one cannot obtain Eq.~\eqref{eq:msd_multi} via the simple rescaling of Eq.~\eqref{eq:lambdas} due to the fact that products of $ \Gamma $ matrices necessarily appear, and they will generically contain an anti-symmetric part (which cannot be converted in convergence and shear-like terms)\footnote{One could argue that $ \Gamma_{\mathrm{c}m} $ can just be reabsorbed in the main lens model displacement angle $ \vec\alpha $, but here we keep an explicit $ \Gamma_{\mathrm{c}m}  $ term in light of the physical interpretation of such mass sheets, which can come from internal cores (like the ultralight dark matter examples shown in Ref.~\cite{Blum:2021oxj,Blum:2024igb}) or from the influence of nearby group (see e.g. Ref.~\cite{Teodori:2023nrz}). Both cases can have non-completely negligible $ \Gamma_{\mathrm{c}m} $ which do not come from the main lens model (usually assumed to be an elliptical power law or a composite stars plus Navarro-Frenk-White (NFW) profile). Moreover, the explicit inclusion of $ \Gamma_{\mathrm{c} m} $ makes the connection between internal and external sheet contributions more apparent.}.. However, we can circumvent this issue if we consider $ \kappa_{\mathrm{c} m} $ and $ \Gamma_{\mathrm{c}m} $ of the same order of external convergences and shears, meaning we retain terms only up to first order in these quantities (consistently with the tidal approximation we used to write Eq.~\eqref{eq:multiplane}). We thus compute the $ \lambda $ factors in Eq.~\eqref{eq:lambdas} only up to first order in mass sheet convergences and shears. We return back to what happens when one relaxes this first-order approximation in Sec.~\ref{s:non_linear}, but we anticipate that if $ \Gamma_{\mathrm{c} m}=0 $, the degeneracy we discuss can be made exact . 
%We remark that such approximation is not needed if $ \Gamma_{\mathrm{c}m} =0 $, i.e. for a perfect core mass sheet

\subsection{The rescaling}
At first order in convergences and shears, we can write\footnote{To not complicate too much the notation, in a sum like $ \sum_{i}^{j} $, if $ j< i $, we mean no sum.}
\begin{equation}\label{eq:expand}
\sum_{m=0}^{n-1} \qty[\mathbb{I}-\kappa(z_n,z_m) - \Gamma(z_n,z_m)] C_{nm} (\kappa_{\mathrm{c}m} + \Gamma_{\mathrm{c}m})\vec\beta_m \simeq 
\sum_{m=0}^{n-1} C_{nm} (\kappa_{\mathrm{c}m} + \Gamma_{\mathrm{c}m}) \qty(\vec\theta - \sum_{l=0}^{m-1} C_{ml} \vec{\alpha}_l) \ ;
\end{equation}
focusing on the parts proportional to the displacement angles, we have (swapping the $ m \leftrightarrow l $ mute indices for later convenience)
\begin{equation}
-\sum_{l=1}^{n-1} \sum_{m=0}^{l-1}  C_{nl} (\kappa_{\mathrm{c}l} + \Gamma_{\mathrm{c}l})C_{lm} \vec{\alpha}_m = -\sum_{m=0}^{n-2} \sum_{l=m+1}^{n-1}  C_{nl} (\kappa_{\mathrm{c}l} + \Gamma_{\mathrm{c}l})C_{lm} \vec{\alpha}_m \ ;
\end{equation} 
thanks to these manipulations, we can rewrite the original lens equation in the form
\begin{align}
\begin{aligned}
\vec\beta_n &= \qty[\mathbb{I}-\kappa(z_n,0) - \Gamma(z_n,0) - \sum_{m=0}^{n-1} C_{nm}(\kappa_{\mathrm{c} m} +\Gamma_{\mathrm{c} m})  ]\vec\theta \\
&- \sum_{m=0}^{n-1} \qty[\mathbb{I}-\kappa(z_n,z_m) - \Gamma(z_n,z_m) - \sum_{l=m+1}^{n-1}  \frac{C_{nl} C_{lm}}{C_{nm}} (\kappa_{\mathrm{c}l} + \Gamma_{\mathrm{c}l}) ] C_{nm} \vec\alpha_m(\vec\beta_m) 	\ .
\end{aligned}
\end{align}
The rescaling we need can now easily be seen to be
\begin{align} \label{eq:rescalings}
\lambda_n &= \qty(1 - \kappa_n - \sum_{m=0}^{n-1}C_{nm}\kappa_{\mathrm{c} m})^{-1} \ , \\
%\lambda^\alpha_n &= \text{ choice } = \lambda_N \qty(1 - \kappa_{Nn} - \sum_{l=n+1}^{N-1} \frac{C_{Nl} C_{ln}}{C_{Nn}} \kappa_{\mathrm{c}n} ) \ , \\
\lambda_{nm} &= \frac{\lambda_n}{\lambda^\alpha_m} \qty( 1 - \kappa_{nm} - \sum_{l=m+1}^{n-1} \frac{C_{nl} C_{lm}}{C_{nm}} \kappa_{\mathrm{c}l} ) \ ,  \\
\lambda_n^{\rm L} &= \lambda_n^{-1} \ , \\
\tilde\Gamma_{nm} &= \qty(1-\kappa_{nm} - \sum_{l=m+1}^{n-1} \frac{C_{nl} C_{lm}}{C_{nm}}\kappa_{\mathrm{c} l})^{-1} 
\qty(\Gamma_{nm} + \sum_{l=m+1}^{n-1} \frac{C_{nl} C_{lm}}{C_{nm}}\Gamma_{\mathrm{c}l}) \ , \\
 \tilde\Gamma_{n} &= \lambda_n \qty(\Gamma_{n} +\sum_{m=0}^{n-1}C_{nm}\Gamma_{\mathrm{c} m}) \ .
\end{align}
From the last two equations, one can easily infer the values to give to $ \lambda_n^{\Gamma} $ and $ \Gamma^{(\mathrm{c})}_{nm} $, $ \Gamma^{(\mathrm{c})}_{n} $, introduced in Eq.~\eqref{eq:lambdas}.
The only arbitrariness in such rescaling comes from the value to associate to $ \lambda^\alpha_n $. This arbitrariness is crucial, as it will imply that some overall convergence factors will remain completely unconstrained in the multiple plane lens equation. We will see that, given our conventions on the normalization of $ \vec{\alpha}_n $, there is a natural choice for $ \lambda^\alpha_n $. But first, we need to better understand the structure of this seemingly convoluted rescaling structure. Luckily, it turns out that this scaling relation assumes a very simple and natural form.

%Notice that, if redshift and cosmological model uncertainties are deemed to be small (enough), then the angular diameter distance expression $ C_{nm} $ is determined, hence $ \lambda_{nm} = \tilde{C}_{nm}/C_{nm} $ is in principle measurable\footnote{Uncertainty on $ H_0 $ does not play a factor, as dependence on it factors out in the ratio.}. We can find a more intuitive form for all these $ \lambda $ factors.
\subsection{Internal and external sheets on the same footing}
Recall (see e.g.~\cite{Bartelmann:2010fz,Teodori:2022ltt} and App.~\ref{s:der}) that
the cosmological convergence between comoving distance $ w_{m} $ and $w_n > w_m$ corresponding to redshift $ z_m $ and $ z_n $ respectively, in the direction $\hat n$ on the sky can be written (in flat $ \Lambda $CDM) as
\begin{equation} \label{eq:kappacosmo}
	\kappa_{\rm tot}(z_n, z_m; \hat{n})=\frac{3H^2_0\Omega_{\rm m}}{2}\int_{w_m}^{w_n} \dd{w} \frac{(w_{n}-w)(w - w_{m})}{w_{n} - w_{m}}(1 + z(w)) \delta(\hat n,w)  \ ,
\end{equation}
where $ H_0 $ is the Hubble constant and $ \Omega_{\rm m} $ the matter density parameter today.
Consider the case where we include the internal mass sheets in $ \delta $, that is $\delta = \delta_{\rm m} + \delta_{\rm sheet}$, where $ \delta_{\rm m}  $ is the cosmological contribution giving rise to external convergences $ \kappa(z_n,z_m) $, whereas $ \delta_{\rm sheet} $ is the internal mass sheet contribution localized in the multiple redshift planes. Then, we can write (dropping $ \hat{n} $)
\begin{equation}
	\kappa_{\rm tot}(z_n, z_m) = \kappa(z_n,z_m) +  \frac{3H^2_0\Omega_{\rm m}}{2} \sum_{k=m+1}^{n-1} \frac{(w_{n}-w_k)(w_k - w_{m})}{w_{n} - w_{m}}(1 + z_k)^2  \underbrace{\int \dd{x_{\rm pr}}\frac{\rho( x_{\rm pr}, z_k) }{\rho_{\rm m}(z_k)} }_{=:\Sigma (z_k) / \rho_{\rm m}(z_k)} \ .
\end{equation} 
$x_{\rm pr}$ is the proper transverse distance at redshift $z_k$, $\rho_{\rm m}(z_k)$ is the matter background density at redshift $z_k$,  
\begin{equation}
	\rho_{\rm m}(z_k) = \frac{3H^2_0}{8\pi G} \Omega_{\rm m} (1+z_k)^3 \ ,
\end{equation}
and $ \Sigma (z_k) $ is the column density sitting at redshift $ z_k $.
We can thus write the contributions of internal mass sheets $ \kappa_{\mathrm{c} m} $ to a certain external convergence quantity as the following sum, recalling the flat-space FRWL expression for angular diameter distances $ D_{\rm A}(z_n,z_m) = (w_n - w_m)/(1+z_n) $,
\begin{empheq}{align}
	\kappa_{\rm tot}(z_n, z_m)= \kappa(z_n,z_m) + 4\pi G \sum_{l=m+1}^{n-1} \frac{D_{\rm A}(z_n,z_l) D_{\rm A}(z_l, z_m) }{D_{\rm A}(z_n,z_m)} \Sigma (z_l) \ .
\end{empheq}
We will call this quantity corrected external convergence, where both internal and external sheets are placed on the same footing.

Notice that the choice Eq.~\eqref{eq:choice} corresponds to the critical density definition
\begin{equation} \label{eq:sigmac}
	\Sigma_{\rm crit}(z_l) = \frac{D_{\rm A}(z_N,0)}{4\pi GD_{\rm A}(z_N, z_l) D_{\rm A}(z_l,0) } \implies \kappa_{\mathrm{c} l} = 4\pi G \frac{D_{\rm A}(z_N, z_l) D_{\rm A}(z_l,0) }{D_{\rm A}(z_N,0)} \Sigma (z_l)\ ,
\end{equation}
from which we can finally write
\begin{empheq}[box=\fbox]{align} \label{eq:kappatot}
	\kappa_{\rm tot}(z_n, z_m)= \kappa(z_n,z_m) +  \sum_{l=m+1}^{n-1} \frac{C_{nl} C_{lm}}{C_{nm}} \kappa_{\mathrm{c} l} \ , \  \kappa_{\rm tot}(z_n, 0)= \kappa(z_n,0) +  \sum_{l=0}^{n-1} C_{nl} \kappa_{\mathrm{c} l} \ , 
\end{empheq}
where we exploited
\begin{equation}
\frac{C_{nl} C_{lm}}{C_{nm}} = C_{nl} \text{ if } w_m =0 \ .	
\end{equation}
Same reasoning applies for shear matrices, 
\begin{equation} \label{eq:gammatot}
	\Gamma_{\rm tot}(z_n, z_m)= \Gamma(z_n,z_m) +  \sum_l \frac{C_{nl} C_{lm}}{C_{nm}} \Gamma_{\mathrm{c} l} \ , \
	\Gamma_{\rm tot}(z_n, 0)= \Gamma(z_n,0) +  \sum_l C_{nl} \Gamma_{\mathrm{c} l} \ . 
\end{equation}
We thus see that the rescaled parameters assume a much simpler, more transparent (and somewhat expected) form,
\begin{align} \label{eq:rescalings2}
	\lambda_n &= \frac{1}{1-\kappa_{\rm tot}(z_n, 0)} \ , \\
	%\lambda^\alpha_n &= \text{ choice } = \lambda_N \qty(1 - \kappa_{Nn} - \sum_{l=n+1}^{N-1} \frac{C_{Nl} C_{ln}}{C_{Nn}} \kappa_{\mathrm{c}n} ) \ , \\
	\lambda_{nm} &= \frac{1}{\lambda^\alpha_m} \frac{1-\kappa_{\rm tot}(z_n, z_m)}{1-\kappa_{\rm tot}(z_n, 0)} \ ,  \\
	\lambda_n^{\rm L} &= 1-\kappa_{\rm tot}(z_n, 0) \ , \\
\label{eq:gamma_nm}	\tilde\Gamma_{nm} &= \frac{\Gamma_{\rm tot}(z_n, z_m)}{1-\kappa_{\rm tot}(z_n, z_m)} \ , \\
\label{eq:gamma_n}	\tilde\Gamma_{n} &= \frac{\Gamma_{\rm tot}(z_n, 0)}{1-\kappa_{\rm tot}(z_n, 0)} \ .
\end{align}
Notice that the rescaling of shears relates to a generalization of reduced shears. 

\subsection{Effective angular diameter distances}
We can simplify things even further, defining
effective angular diameter distances
\begin{empheq}[box=\fbox]{align} \label{eq:Deff_def}
D^{\rm eff}_{\rm A}(z_{n},z_m) = (1 - \kappa_{\rm tot}(z_n, z_m)) D_{\rm A}(z_{n},z_m) \ .
\end{empheq}
The original lens equation can be then written as
\begin{equation} \label{eq:Deff}
D^{\rm eff}_{\rm A}(z_{n},0)\vec{\tilde\beta}_n = (\mathbb{I} - \tilde\Gamma_n)D^{\rm eff}_{\rm A}(z_{n},0)\vec{\theta} - \sum_{m=0}^{n-1} (\mathbb{I} - \tilde{\Gamma}_{nm}) D^{\rm eff}_{\rm A}(z_{n},z_m) \lambda_n^\alpha\hat{\alpha}_m( D^{\rm eff}_{\rm A}(z_m,0)\vec{\tilde\beta}_m) \ ,
\end{equation}
which has the same form of the MSD-reduced equation. 
Once we better understand the role of $ \lambda_n^\alpha $, everything becomes straightforward, as one can pass from the original lens equation to the MSD-reduced one by simply replacing $ D_{\rm A} $ with $ D^{\rm eff}_{\rm A} $ and using ``tilde quantities'', like corrected external reduced shears instead of the ``true'' shear values.
This will allow us to immediately find the time delay inference of the MSD-reduced model in Sec.~\ref{s:time_delay}.

To understand what becomes observable and what is still unobservable in the multiple plane lens equation, it will be useful to rewrite Eq.~\eqref{eq:Deff} as
\begin{equation} 
	\vec{\tilde\beta}_n = (\mathbb{I} - \tilde\Gamma_n)\vec{\theta} - \sum_{m=0}^{n-1} (\mathbb{I} - \tilde{\Gamma}_{nm}) \frac{D^{\rm eff}_{\rm A}(z_{n},z_m)}{D^{\rm eff}_{\rm A}(z_{n},0)} \lambda_n^\alpha\hat\alpha_m( D^{\rm eff}_{\rm A}(z_m,0)\vec{\tilde\beta}_m) \ ;
\end{equation}
we see that the ratios $ D^{\rm eff}_{\rm A}(z_{n},z_m)/D^{\rm eff}_{\rm A}(z_{n},0)  $ are degenerate with the $ \lambda_n^{\alpha} $ factors. However, there are (at most) $ N $ $ \lambda_n^{\rm \alpha} $ factors, whereas there are $ N(N+1)/2 $ angular diameter distance ratios in the form $ D^{\rm eff}_{\rm A}(z_{n},z_m)/D^{\rm eff}_{\rm A}(z_{n},0) $. Meaning, one can choose the $ \lambda_n^{\alpha} $ factors to reabsorb $ N $ angular diameter distance ratios of one's choice, and there are remaining $ N(N-1)/2 $ angular diameter distance ratios which cannot be reabsorbed and are indeed observable.

To be concrete, suppose we absorb in the $ \hat{\alpha}_m  $ normalizations the following factors, via the following choice of $ \lambda^\alpha_n $ 
\begin{equation}\label{eq:lambda_alpha}
	\lambda^\alpha_n  =  \frac{1-\kappa_{\rm tot}(z_N, z_n)}{1-\kappa_{\rm tot}(z_N, 0)} \ ;
\end{equation}
such a choice for $ \lambda^\alpha_n $ is inspired by the choice Eq.~\eqref{eq:choice} for the displacement angle expression. Then, we can define
\begin{equation}\label{eq:ang_ratio}
	\tilde{C}_{nm} := C_{nm} \frac{1-\kappa_{\rm tot}(z_{N},0)}{1-\kappa_{\rm tot}(z_{n},0)}\frac{1-\kappa_{\rm tot}(z_n, z_m)}{1-\kappa_{\rm tot}(z_{N}, z_m)} = \frac{D^{\rm eff}_{\rm A}(z_{N},0)  D^{\rm eff}_{\rm A} (z_n, z_m)}{ D^{\rm eff}_{\rm A}(z_{n},0)  D^{\rm eff}_{\rm A}(z_{N}, z_m)} \ ,
\end{equation}
which, at first order in convergences, can be rewritten as
\begin{equation}\label{eq:diffkappa}
	\tilde{C}_{nm} \simeq C_{nm}  \qty{1-(\kappa_{\rm tot}(z_n, z_m) - \kappa_{\rm tot}(z_N, z_m)) + (\kappa_{\rm tot}(z_n, 0) - \kappa_{\rm tot}(z_N, 0))} =: C_{nm}(1- \Delta\kappa^{\rm tot}_{Nnm}) \ .
\end{equation}
If $ H_0 $ is the only cosmological parameter one cares about, $\Delta\kappa^{\rm tot}_{Nnm} $ terms are in principle observable, meaning that all the $ \tilde{C}_{nm} $ factors are observable (if redshift and cosmological model uncertainties are deemed to be small enough for a good determination of $ D_{\rm A}(z_n,z_m) $ using redshifts alone\footnote{Knowledge of $ H_0 $ is not needed, as it cancels in the $ C_{nm} $ ratio.}). Notice that, since $ \tilde{C}_{Nm} /C_{Nm}= 1$ by definition, $ \tilde{C}_{nm} $ factors contain $ N(N-1)/2 $ non-trivial factors, in agreement with the number of observable angular diameter distance ratios in an $ N $ multiple plane system. Notice that $\Delta\kappa^{\rm tot}_{Nnm} $ terms contain a mixture of differential external convergences and differential internal sheet convergences. 

However, if one wants to determine other cosmological parameters like $ \Omega_{\rm m} $, as noted in Ref.~\cite{Johnson:2025lhy} (which however focuses only on cosmological convergences), measurements of angular diameter distance ratios on multiple plane systems, based on inference of $ \tilde{C}_{nm} $, are biased by a factor
\begin{equation} \label{eq:bias}
\frac{\tilde{C}_{nm}}{C_{nm}} \simeq 1 - \Delta\kappa^{\rm tot}_{Nnm} \ ,
\end{equation}
if convergence on the LOS are not taken into account. Hence, measurements of cosmological parameters like $ \Omega_{\rm m} $, $ \Omega_\Lambda $ are biased by $ 1 - \Delta\kappa^{\rm tot}_{Nnm}  $.
Our further comment to Ref.~\cite{Johnson:2025lhy} concerns the contribution to biases on angular diameter distance ratios determination coming from internal mass sheets as well.

We can thus finally write the lens equation, valid both for the original model and the MSD-reduced model, in the simple form
\begin{empheq}[box=\fbox]{align} \label{eq:lens_tilde}
	\vec{\tilde\beta}_n = (\mathbb{I} - \tilde\Gamma_n)\vec{\theta} - \sum_{m=0}^{n-1} (\mathbb{I} - \tilde{\Gamma}_{nm}) \tilde{C}_{nm}\vec{\tilde\alpha}_m( D^{\rm eff}_{\rm A}(z_m,0)\vec{\tilde\beta}_m) \ .
\end{empheq}
To obtain this, one can take the original model, remove all the mass sheet terms, and substitute angular diameter distances with effective angular diameter distances of Eq.~\eqref{eq:Deff_def} (together with ``tilde'' quantities for displacement angles $ \vec{\alpha}_m $, shears $ \Gamma_m $ and source angles $ \beta_m $).

We remark that the choice Eq.~\eqref{eq:lambda_alpha} for $ \lambda_n^\alpha $ is by no means mandatory; one can decide to reabsorb any $ N $ different angular diameter distance ratios from the $ N(N+1)/2 $ available. In the end, the observables will always be some combination of the $\Delta\kappa^{\rm tot}_{lnm}$ factors we defined, which encompass all the possible convergence differences between two planes (which, in a $ N $ plane system, there are $ N(N-1)/2 $ independent ones).  

%We see that what becomes observable in multiple source systems are ratios of effective angular diameter distances $ D^{\rm eff}_{\rm A}(z_{n},z_m)/ D^{\rm eff}_{\rm A}(z_{n},0) $. Equivalently, if redshift and cosmological model uncertainties are deemed to be small (enough), what becomes observable are the differential convergences
%\begin{equation}\label{eq:diffkappa}
%\frac{1 - \kappa_{\rm tot}(z_n, z_m)}{1 - \kappa_{\rm tot}(z_n, 0)} \frac{1 - \kappa_{\rm tot}(z_l, 0)}{1 - \kappa_{\rm tot}(z_l, z_m)} \simeq 1-(\kappa_{\rm tot}(z_n, z_m) - \kappa_{\rm tot}(z_l, z_m)) + (\kappa_{\rm tot}(z_n, 0) - \kappa_{\rm tot}(z_l, 0)) =: 1 -\Delta\kappa^{\rm tot}_{lnm} \ .
%\end{equation}
%The newly defined $\Delta\kappa^{\rm tot}_{lnm} $ is an observable, as one can absorb in the deflector lens model $ \hat{\alpha}_m $ (via what we called $ \lambda_m^\alpha $) in the MSD-transformed model at most one $ (1 - \kappa_{\rm tot}(z_n, z_m))/(1 - \kappa_{\rm tot}(z_n, 0)) $ factor for fixed $ m $ and varying $ n $, hence ratios of these factors must impact the lens equation even in the MSD model Eq.~\eqref{eq:msd_multi}.

%Explicitly,
%\begin{equation}
%\Delta\kappa^{\rm tot}_{Nmn} = 
%\end{equation}

We can conclude that, as noted in Ref.~\cite{Teodori:2022ltt}, the multiple plane lens equation is oblivious to some overall corrected external convergence constants, it is sensitive only to certain differences $ \Delta\kappa^{\rm tot}_{Nnm} $. The main difference with Ref.~\cite{Teodori:2022ltt} is that there we did not account for lens-lens coupling, meaning there is only one deflector and $ N$ angular diameter distance ratios. Having the freedom to rescale away only one angular diameter distance ratio, one ends up with one overall convergence factor which is completely unconstrained and $ N-1 $ observable $ \Delta \kappa^{\rm tot}_{Nn0} $ factors. 

A possible help in breaking the MSD degeneracy as presented here comes from stellar kinematics, which can constrain how much the $ \lambda_n^\alpha $ factors can really absorb. This is possible as stellar kinematics can give a prior on the mass of the lens. This is completely analogous to the situation one has in single plane lens modeling. We return back to this point in Sec.~\ref{s:cluster}.

%To summarize, the multiple plane MSD corresponds to the rescaling Eq.~\eqref{eq:lambdas},\eqref{eq:rescalings}, where the only (in principle) observable effect comes from the effective angular diameter distances ratio, which can be translated in differential convergences, see Eq.~\eqref{eq:ang_ratio}.
%To connect this last result with the findings of~\cite{Teodori:2022ltt}, we can rewrite Eq.~\eqref{eq:ang_ratio} at first order in convergences
%\begin{equation}
%\tilde{C}_{nm} \simeq C_{nm} \ .
%\end{equation}
%We again see that internal sheet effects can become distinguishable from differential external convergences if the former are bigger than what it is cosmologically expected from the latter. In other words, looking at
%\begin{equation}
%\frac{\tilde{C}_{nm}}{C_{nm}} \simeq ... \ ,
%\end{equation}
%if the measured LHS of the previous is too large to be explained by the cosmological differential convergences alone, then one can infer the presence of an internal mass sheet (or better, mis-modeling of the lens profile).

\section{Time delays in the MSD-reduced model} \label{s:time_delay}
As we show in App.~\ref{s:tdelay} and as shown in e.g. Refs.~\cite{Schneider:1992,Schneider:2014vka,McCully:2016yfe,Fleury:2021tke}, the time delay expression for multiple plane systems in the presence of LOS effects reads
\begin{equation}
	\Delta T(\{\vec{\beta}_m\}, \beta_N) =\sum_{n=0}^{N-1}  D_{\rm dt}^n \qty( \frac{\vec{B}^\top_n A^{\rm tot}_{n+1,n}(A^{\rm tot}_{n+1})^{-1} A^{\rm tot}_{n}\vec{B}_n}{2C_{n+1,n}} - \Psi_n(\vec{\beta}_n) ) \ , \ D_{\rm dt}^n:= (1+z_n) \frac{D_{\rm A}(z_N,0) D_{\rm A}(z_n,0)}{D_{\rm A}(z_N, z_n)} \ ,
\end{equation}
where
\begin{equation}
A^{\rm tot}_{nm} := (1 - \kappa_{\rm tot}(z_n, z_m))\mathbb{I} - \Gamma_{\rm tot}(z_n,z_m) \ ,  \ \vec{B}_n :=  (A^{\rm tot}_{n+1,n})^{-1} ( A^{\rm tot}_{n+1}(A^{\rm tot}_n)^{-1} \vec\beta_n - \vec\beta_{n+1}) 
\end{equation}
(see App.~\ref{s:tdelay} for further details).
Notice that the lensing potential $ \Psi $, in the MSD-reduced model, becomes
\begin{equation}
\tilde\Psi_n(\vec{\tilde{\beta}}) = \frac{\lambda_n^\alpha}{\lambda^{\rm L}_n} \Psi_n(\lambda_n\vec{\tilde\beta}) \ .
\end{equation}

With our insight that we can absorb all convergence factors in the redefinition of angular diameter distances, the MSD-reduced time delay immediately reads
\begin{empheq}[box=\fbox]{align} \label{eq:deltaT}
	\Delta T(\{\vec{\tilde\beta}_m\}, \vec{\tilde\beta}_N) =\sum_{n=0}^{N-1}  \tilde{D}_{\rm dt}^n \qty( \frac{\vec{\tilde{B}}^\top_n \tilde{A}_{n+1,n}\tilde{A}^{-1}_{n+1} \tilde{A}_{n}\vec{\tilde{B}}_n}{2\tilde{C}_{n+1,n}} - \tilde{\Psi}_n(\vec{\tilde\beta}_n) ) \ , \ \tilde{D}_{\rm dt}^n = (1+z_n) \frac{D^{\rm eff}_{\rm A}(z_N,0) D^{\rm eff}_{\rm A}(z_n,0)}{D^{\rm eff}_{\rm A}(z_N, z_n)} \ ,
\end{empheq}
where
\begin{equation}
\tilde{A}_{nm} = \mathbb{I} - \tilde{\Gamma}_{nm} \ , \ \vec{\tilde{B}}_n = \tilde{A}_{n+1,n}^{-1} ( \tilde{A}_{n+1}\tilde{A}_n^{-1} \vec{\tilde{\beta}}_n - \vec{\tilde{\beta}}_{n+1})  \ .
\end{equation}
The way it is written, Eq.~\eqref{eq:deltaT} is valid for both the original model and the MSD-reduced model. The difference between the two is the following: an hypothetical observer aware of the corrected external convergence factors will write $ \tilde{D}_{\rm dt}^n $ using effective angular diameter distances with the convergence factors explicit, whereas the observer which cannot access such information (the MSD-reduced observer, that is us) will have to deal somehow with the uncertainties coming from the $ \kappa^{\rm tot} $ factors. Depending on how the latter is done, one risks to have biased inference on the main cosmological parameters time delays can measure, i.e. $ H_0 $.

%Now, there are at most $ N $ time-delay equations one can use\footnote{This is a very optimistic scenario, even in the $ N=2 $ case, since we would need reliable time-varying sources in all planes to have $ N $ different measurements of time delays.}, and only $ N(N-1)/2 $ convergence differences can be determined from the lens equation. 
Interestingly, $ D_{\rm dt}^n $ terms obey the so-called unfolding relations~\cite{Schneider:1992,Fleury:2020cal}\footnote{We thank D. Johnson and P. Fleury for an assessment on this point. See also Ref.~\cite{Johnson:2026uhh}.}. In our notation, it reads
\begin{equation}
	\frac{C_{ki}}{D_{\rm dt}^i} = \frac{C_{ji}}{D_{\rm dt}^i} + \frac{C_{kj}}{D_{\rm dt}^j} \  \forall i < j< k \ ; 
\end{equation}
this would be just a relation on angular diameter distances, but Ref.~\cite{Fleury:2020cal} showed that it works even when weak lensing effects are included in $ D_{\rm dt}^i $. In our language, we can thus also write
\begin{equation} \label{eq:unfolding}
	\frac{\tilde{C}_{ki}}{\tilde{D}_{\rm dt}^i} = \frac{\tilde{C}_{ji}}{\tilde{D}_{\rm dt}^i} + \frac{\tilde{C}_{kj}}{\tilde{D}_{\rm dt}^j} \quad \forall i < j< k \ ;
\end{equation}
where we seemingly obtain a relation between both internal and external sheets on the different planes. Unfortunately (although as one would expect), Eq.~\eqref{eq:unfolding} relates only the different weak lensing convergences, and no relation is imposed on $ \kappa_{\mathrm{c}m } $. It is easy to show this for the $ N=2 $ case, but for the doubtful reader, we demonstrate it in App.~\ref{s:unfolding}.

However, Eq.~\eqref{eq:unfolding} can be rewritten in the useful form (for $ i<j $)
\begin{equation}\label{eq:unfold_rel}
\frac{\tilde{D}^i_{\rm dt}}{\tilde{D}^j_{\rm dt}} = 1 - \tilde{C}_{ji} \ ;
\end{equation}
this has the important implication that, with knowledge of $ \tilde{C}_{ji} $ and e.g. $ \tilde{D}^0_{\rm dt} $, one can obtain all the other $ \tilde{D}_{\rm dt}^n $ factors. This is important, as it is very unlikely, even in the $ N=2 $ case, to have reliable time-varying sources in all planes, which would allow measurements of the $ N $ time delays. But thanks to Eq.~\eqref{eq:unfold_rel}, one does not need that. In fact, Eq.~\eqref{eq:unfold_rel} implies that one needs only the multiple plane lens equation Eq.~\eqref{eq:lens_tilde} and one out of the $ N $ time delay equations of Eq.~\eqref{eq:deltaT} to predict the other $ N-1 $ time delays. The clear advantage of this fact is that one needs only one time-variable source among all the available planes to obtain expressions for all $ \tilde{D}_{\rm dt}^n $ (assuming accurate lens modeling inference for $ \tilde{C}_{nm} $ is possible); the disadvantage is that, in the lucky scenario where one has multiple time delay measurements, the extra time delay measurements would not give additional constraints, but only (important) consistency checks. 

%Crucially, the convergence factors entering $ \tilde{D}^n_{\rm dt} $ cannot be expressed as combinations of $ \Delta\kappa_{Nnm} $, as angular diameter distance ratios in $ \tilde{D}^n_{\rm dt} $ have a different structure with respect to the ones in $ \tilde{C}_{nm} $ (in particular, an odd number of $ D_{\rm A}^{\rm eff} $ versus an even number of $ D_{\rm A}^{\rm eff} $ terms), since %falling short of the $ N(N+1)/2 +1$ convergence factor in the problem.
%\begin{equation}
%\frac{\tilde{D}^n_{\rm dt}}{D^n_{\rm dt}} \simeq 1 - \kappa^{\rm tot}_{N} - \kappa^{\rm tot}_n + \kappa^{\rm tot}_{Nn} = 1 - \Delta\kappa^{\rm tot}_{Nln} + \kappa^{\rm tot}_{ln} - \kappa^{\rm tot}_l-\kappa^{\rm tot}_n \ ,
%\end{equation}
%for $ l>n $, %(hence it will not work if $ n = N-1 $), 
%and there will be always one differential convergence factor hanging (for reference, there are $ (N+1)(N+2)/2 $ corrected external convergence factors). In other words, $ \tilde{D}^n_{\rm dt} $ are yet $ N $ unknowns. 
Even with knowledge of all the $  \tilde{D}^n_{\rm dt} $, one cannot recover $ H_0 $, due to lack of another independent equation which is not there. To conclude, even with knowledge of all $ \tilde{D}^n_{\rm dt} $, there is still one factor which mixes $ H_0 $ and the convergence factors in the $ \tilde{D}^n_{\rm dt}/D^n_{\rm dt} $ terms. In a $ N+1 $ parameter space composed of $ H_0 $ and $\tilde{D}^n_{\rm dt}  $, %(the latter obtained via the time delay equation), 
the degeneracy lives on a $ 1 $-D line in this $ N+1 $ dimensions parameter space, meaning the MSD is still retained, and the $ H_0 $ measurement still suffers from the degeneracy.
%Still, there are new corrected external convergence factors that become observable independently of the $ H_0 $ knowledge (but still assuming other cosmological parameters are known), that is
%\begin{equation} \label{eq:Ddt_obs}
%\frac{\tilde{D}^n_{\rm dt}}{D^n_{\rm dt}} \frac{D^m_{\rm dt}}{\tilde{D}^m_{\rm dt}} \simeq 1- \kappa_n + \kappa_m + \kappa_{Nn} - \kappa_{Nm} =: 1 - \Delta\kappa^{D_{\rm dt}}_{Nnm} \ .
%\end{equation}
%In particular, there are $ N-1 $ new independent differential corrected external convergences $ \Delta\kappa^{D_{\rm dt}}_{Nn0} $, where $ \Delta\kappa^{D_{\rm dt}}_{Nnm} = \Delta\kappa^{D_{\rm dt}}_{Nn0} - \Delta\kappa^{D_{\rm dt}}_{Nm0} $.

%To make this concrete, suppose one declares $ H_0  $ from the first source time delay $ \Delta T (\vec{\beta}_1 \vec{\beta}_0) $, hence the $ H_0 $ it will get will be biased by
%\begin{equation}
%\frac{1}{\tilde{H}_0} = \frac{1-\kappa_N - \kappa_0 + \kappa_{N0}}{H_0} \ ;
%\end{equation}
%then, the only way he can use the other time delays is just to constrain yet other 

%We see that multiple source systems do not break this approximate MSD degeneracy, as the ``monopole'' factors of all the convergences in the lensing problem is completely unconstrained. 
Interestingly, the LOS part of the $ \Delta\kappa^{\rm tot}_{Nnm} $ quantities would not be allowed to be any value one wants, hence the presence of internal sheets in the different lens planes may be made manifest on unexpectedly large observed values of $ \Delta\kappa^{\rm tot}_{Nnm} $, large with respect to the expected cosmological contribution to Eq.~\eqref{eq:diffkappa}. This was a main point of Ref.~\cite{Teodori:2022ltt}, which we here generalize to the case where lens-lens coupling must be taken into account.%, and when one has multiple time delays available.  

\section{An example of a multiple source system non-breaking the MSD} \label{s:cluster}
We can make an explicit example illustrating, on the best-case scenario where one can have a good determination of lens model and time delay surfaces on all the lens planes, how the multiple source MSD would work. For our example, we use the system J1721+8842 identified in Ref.~\cite{Dux:2024vvq}, there renamed the ``zigzag'' lens, which is a double source system with $ z_0 =0.184 $, $ z_1 = 1.885 $, $ z_2= 2.38 $. This is by no means the only non-cluster multiple source system known (see e.g.~\cite{Gavazzi:2008aq,Collett:2020lii,TU:2009cyy,Tanaka:2016pdb,Schuldt:2019vza,Bolamperti_2023}), we take it here just as an example. We can consider a setting where we associate ``typical'' external convergence values to such a system, using the estimates of Ref.~\cite{Teodori:2022ltt}\footnote{Details on how these estimates are computed can be found in App. B of Ref.~\cite{Teodori:2022ltt}, we here note that we used a standard $ \Lambda $CDM cosmological model with the parameters of Ref.~\cite{Akrami:2018vks}. The non-linear power spectrum, needed to get more reliable estimates of LOS convergences, is obtained from  HALOFIT~\cite{Peacock:2014,Takahashi:2012em}. A notebook with the implementation of the computation of such estimates can be found \href{https://github.com/lucateo/Comments_MSD/blob/main/Notebooks/delta_kappa_nonlinear.ipynb}{here}.}. Moreover, we add two generic core factors on the different lens planes, $ \kappa_{\mathrm{c} 0} = 0.05 $ and $ \kappa_{\mathrm{c} 1} = 0.03 $, see Table~\ref{tab:mock}. We now consider different aspects of this scenario.

\paragraph{Stellar kinematics.} 
First, we comment on how stellar kinematics can constrain the MSD and eventually the $ \kappa_{\mathrm{c} n} $ terms. With $ \tilde\sigma_{\mathrm{los},n} $ as the velocity dispersion of the $ n $-th lens without LOS and internal sheet effects (that is, the lens model originating the displacement angle $ \vec{\tilde\alpha}_{n} $) and $ \sigma_{\mathrm{los}, n} $ the lens with both LOS and internal sheets taken into account (that is, the model originating the displacement angle $ \vec\alpha_n + (\kappa_{\mathrm{c} n} + \Gamma_{\mathrm{c}n})\vec\theta $),
\begin{equation} \label{eq:kin}
	\frac{\sigma^2_{\mathrm{los} ,n}}{\tilde\sigma^2_{\mathrm{los},n}} = \frac{1}{\lambda^\alpha_n}+ \delta_{\rm c} \ , \ \delta_{\rm c} := \frac{\sigma^2_{\mathrm{c} ,n}}{\tilde\sigma^2_{\mathrm{los},n}}  \ , 
\end{equation}
where $ \sigma_{\mathrm{c} ,n} $ is the velocity dispersion contribution of the internal core. See App.~\ref{s:stellar_kin} for more details. The factor $ 1/\lambda^\alpha_n $, Eq.~\eqref{eq:lambda_alpha}, comes from the different normalization between $ \vec\alpha_n $ and $ \vec{\tilde{\alpha}}_n $. In the perfect internal mass sheet limit, $ \delta_{\rm c}=0 $, but finite core effects would make $ \delta_{\rm c} $ different from zero. In particular, notice that in the case where $ \lambda_n^{\alpha}>1 $ (which, for the single lens plane case, would imply a upward bias on the $ H_0 $ inference from time delays), a non-zero $ \delta_{\rm c} $ would make the stellar kinematics constraint of the MSD \emph{weaker}%\footnote{The case $ \lambda^\alpha_n < 1  $ would physically correspond to a void, and one needs a specified model for $ \delta_{\rm c} $ to understand whether constraints become weaker or stronger in that case. Whereas for cores, $ \delta_{\rm c} \ge 0$ for any reasonable model (with density monotonically decreasing), making the constraint always weaker.}
, highlighting the importance of modeling internal mass sheets with finite-core parametrization, differently from what it is done on the analyses of Refs.~\cite{Birrer:2020tax,TDCOSMO:2025dmr}\footnote{The conclusion of Ref.~\cite{Birrer:2020tax} that these finite-core effects are negligible were unfortunately affected by a small bug, in particular they did not allow for a changing stellar distribution effective radius $ r_{\rm eff} $ in their pipeline.}. See comment in Ref.~\cite{Blum:2021oxj} for details and an explicit example. 
Even neglecting $ \delta_{\rm c} $, the convergence factors of Table~\ref{tab:mock} would affect stellar kinematics measurements $ \sigma_{\mathrm{los}, n} $ on lens $ 0 $ by at most $ 4 \% $ and lens $ 1 $ by at most $ 6 \% $, following 
\begin{equation} \label{eq:sigma}
\abs{1- \frac{\sigma_{\mathrm{los},n}}{\tilde\sigma_{\mathrm{los},n}}} \simeq 0.5\abs{\kappa_{Nn} - \kappa_{N}} \ ;
\end{equation}
hence, stellar kinematics in this example would play a completely analogous role to single plane lens settings.
%\footnote{At most, because finite core effects of internal sheets whose central convergence is $ \kappa_{\mathrm{c} l} $ would reduce stellar kinematics influence with respect to an exact mass sheet. See comment in Ref.~\cite{Blum:2021oxj}.}.
%The $ \kappa_{\mathrm{c} m} $ terms come from mismodelling of the lens $ m $. Contributions to $ \kappa_{\mathrm{c}m} $ may come from main profile normalization adjustments~\cite{Shajib:2020ptb}, contributions from nearby groups~\cite{Teodori:2023nrz} or internal dark matter core components~\cite{Blum:2021oxj,Blum:2024igb}.

\paragraph{Time delays.} Assuming that time delays are measured from both the first and second source or/and that knowledge of one time delay plus good inference of the lens model allows the use of the unfolding relation in Eq.~\eqref{eq:unfold_rel}, we can infer $ \tilde{D}_{\rm dt}^0 $ and $ \tilde{D}_{\rm dt}^1 $. The knowledge of both time delay surfaces allows us to put on a line the relation between $ \tilde{D}_{\rm dt}^0/D_{\rm dt}^0 $, $  \tilde{D}_{\rm dt}^1/D_{\rm dt}^1 $ and $ H_0 $, where, within our normalizations,
\begin{equation} \label{eq:bla}
	\frac{\tilde{D}^n_{\rm dt}}{D^n_{\rm dt}} = \frac{(1-\kappa^{\rm tot}_N) (1-\kappa^{\rm tot}_n)}{(1-\kappa^{\rm tot}_{Nn})} \ .
\end{equation}
We show such an exercise in Figure~\ref{fig:example}.
This is useful, as stellar kinematics predicts different prior range for Eq.~\eqref{eq:bla} for different  $ n $, see below.

As an aside, one could worry that different parametrizations (or equivalently, different choices of $ \lambda_n^\alpha $) could change the relation we show in Figure~\ref{fig:example}.
However, with different parametrizations, one can write a different relation between convergence factors and effective time delay distance factors. For example, normalize the displacement angles Eq.~\eqref{eq:choice} with the following alternative choice,
\begin{equation}
	\vec{\alpha}^{\rm alt}_n = \frac{D^{\rm eff}_{\rm A}(z_{n+1}, z_n)}{D^{\rm eff}_{\rm A}(z_{n+1},0)} \hat{\alpha}_n \implies 
	\tilde{C}^{\rm alt}_{nm} = \frac{D^{\rm eff}_{\rm A}(z_{m+1},0)  D^{\rm eff}_{\rm A} (z_n, z_m)}{ D^{\rm eff}_{\rm A}(z_{n},0)  D^{\rm eff}_{\rm A}(z_{m+1}, z_m)} \ ;
\end{equation}
notice that the only non-zero $ \tilde{C}^{\rm alt}_{nm} $ term here is $ \tilde{C}^{\rm alt}_{20} $, which not surprisingly gives rise to the same $ \Delta\kappa_{210} $ observable for the lens system. Instead,
\begin{equation}
	\frac{\tilde{D}^0_{\rm dt,alt}}{D^0_{\rm dt,alt}} = \frac{1 - \kappa^{\rm tot}_1}{1- \kappa^{\rm tot}_{10}}{(1-\kappa^{\rm tot}_{0})}  =  \frac{\tilde{D}^0_{\rm dt}}{D^0_{\rm dt}}\tilde{C}^{\rm alt}_{20} =  \frac{\tilde{D}^0_{\rm dt}}{D^0_{\rm dt}}\frac{1}{\tilde{C}_{10}} \ , \ \frac{\tilde{D}^1_{\rm dt,alt}}{D^1_{\rm dt,alt}} = \frac{\tilde{D}^1_{\rm dt}}{D^1_{\rm dt}} \ .
\end{equation}
As expected, the informative content of the full system obviously does not change with the parametrization, as it is possible to recover the same differential convergence factors with an appropriate use of $ \Delta\kappa_{Nnm} $ factors. %Remind that, within our chosen parametrization, the lens equation can get us $ \Delta\kappa_{Nnm} $. In fact, we can pass from the convergences in the original parametrization to the first source parametrization with

\paragraph{Induced shear.} Realistic internal mass sheets would have an effective shear contribution which changes with radius, coming from finite core effects.
The availability of multiple planes does not change the impact on lens modeling more than what happens on single source systems. %Such effect for typical core parametrization is not more prominent than single plane settings. 
As an example, consider an exponential parametrization of a cored convergence profile
\begin{equation}
\kappa_{\mathrm{c}} (\theta) = \kappa_{\mathrm{c} 0} \e^{-(\theta/\theta_{\rm c})^2} \  \implies \vec\alpha(\vec{\theta}) = \frac{\kappa_{\mathrm{c} 0} \theta^2_{\rm c}}{\theta} \qty( 1 - \e^{-(\theta/\theta_{\rm c})^2}) \frac{\vec{\theta}}{\theta} \ ;
\end{equation}
computing derivatives of the deflection angle, allows us to write the shear as (with $ \vec{\theta} = \theta (\cos \varphi, \sin\varphi) $)
\begin{equation}
(\gamma_1, \gamma_2) = |\gamma| (\cos 2 \varphi, \sin 2 \varphi) \ , \ |\gamma| = \kappa_{\mathrm{c} 0} \qty( \frac{\theta_{\rm c}^2}{\theta^2} - \e^{-(\theta/\theta_{\rm c})^2} \qty(1+\frac{\theta_{\rm c}^2}{\theta^2}) ) \ . 
\end{equation}
In particular, we have the following approximate behaviors, which hold for $ \theta \ll \theta_{\rm c} $
\begin{equation}
|\gamma(\theta \ll \theta_{\rm c})| \sim \kappa_{\mathrm{c} 0} \frac{\theta^2}{\theta_{\rm c}^2} \ ;
\end{equation}
such behavior is typical of cored profiles (see examples shown in Ref.~\cite{Blum:2020mgu}), not only for the exponential example we showed here. In particular, $ \Gamma_{\mathrm{c} l} $ effects are suppressed with respect to the corresponding $ \kappa_{\mathrm{c} l} $ terms by powers of $ \theta/\theta_{\rm c} $. In Table~\ref{tab:mock}, we assumed conservatively that $ \Gamma_{\mathrm{c} l}  $ magnitudes are just a factor of 2 smaller than the respective $ \kappa_{\mathrm{c} l} $ (but assuming the shear to be effectively constant in the region where images form). Still, resulting effective shear $ \tilde{\Gamma} $ corrections from the internal core presence is limited and of the same order of cosmological external convergence/shear, see Table~\ref{tab:mock}. In any case, as pointed out in Ref.~\cite{Etherington:2023yyh}, shear inferred in lens modeling not always can be associated with true cosmological shear, but may instead reflect complexities of the lens environment that are not accounted for in the model (not necessarily coming from unaccounted cores, but also from deviations from ellipticity of the main lens component etc.). This implies that identifying large inferred shear as smoking gun for the presence of unaccounted cores might be problematic. Nevertheless, the possibility of constraining shear differences with the method outlined in Ref.~\cite{Khadka:2024bmw} would still apply, although multiple sources would not introduce anything new as far as this method is concerned.

\paragraph{Constraining the $ H_0 $ degeneracy window.} A possible advantage in the use of multiple sources comes from the possibility of reducing the $ H_0  $-$ \tilde{D}^n_{\rm dt} $ degeneracy line of Figure~\ref{fig:example} via both multiple stellar kinematics priors and cosmological priors; the latter would relate to the expected size of the LOS part of $ \kappa^{\rm tot}(z_n, z_m) $ and correlations between the different sight-lines, together with the relation in Eq.~\eqref{eq:unfolding}. 
Denote $ \delta\kappa^{\rm los}_{nm} $ the expected 1 sigma dispersion of the external convergence part of the corrected convergence factors; these dispersions may come from cosmological priors.
For our example, the value of the (in principle) observable $ |\Delta\kappa^{\rm tot}_{210}| \simeq 0.003 \lesssim |\delta(\Delta\kappa^{\rm los}_{210})| $, meaning that measurement of $ \Delta\kappa^{\rm tot}_{210} $ in our setting will not give any hint on the possible existence of unaccounted internal sheets, despite them being there\footnote{A realistic assessment of whether certain external convergences sizes are indeed unexpected must come from proper system-by-system ray-tracing analyses, on the lines of Refs.~\cite{Hilbert:2008kb,Johnson:2025lhy}. Here we base our claims on the much simpler external convergence variance estimates of Ref.~\cite{Teodori:2022ltt}.}. 
However, this does not need to be the case, see Ref.~\cite{Teodori:2022ltt} for a further assessment on this point.
%However, in our example, $ |\Delta\kappa^{D_{\rm dt}}_{Nnm}| \sim 0.14 $, which may be big with respect to the expected value from external convergences alone One could thus infer the presence of the core from time delay surface ratios as well.
On the stellar kinematics side, assuming a certain $ \delta_\sigma := \Delta \sigma_{\mathrm{los}}/\sigma_{\mathrm{los}} $ precision on the $ \sigma^n_{\rm los} $ measurement, gives the constraint
\begin{equation}
\abs{\kappa_{2} - \kappa_{20}} = \abs{\kappa_1 - \kappa_{10} - \Delta \kappa_{210} } \lesssim 2\delta_\sigma \ , \  \abs{\kappa_{2} - \kappa_{21}} \lesssim 2\delta_\sigma \ ,
\end{equation}
which would roughly translate into
\begin{equation}
\abs{\frac{\tilde{D}_{\rm dt}^n}{D_{\rm dt}^n}} \lesssim 2\delta_\sigma + \abs{\kappa_n} \ .
\end{equation}
A proper treatment of these limits would combine kinematics priors and cosmological priors, but we do not expect our estimates to change in any significant way when translating to a more thorough treatment.
To roughly take such constraints into account, we draw exclusion bands in Figure~\ref{fig:example} to satisfy the previous, with $ \delta_\sigma =0.05 $ and $ \abs{\kappa_0} \sim 0  $, $ \abs{\kappa_1} \lesssim 2\delta_\sigma $.

%In a realistic situation, it would be very hard to have time delays in more than one plane. However, even with only one $ \tilde{D}_{\rm dt}^n/D_{\rm dt}^n $ available, stellar kinematics plus cosmological constraints on the size of the various convergences factors from the multiple planes could help more than what one can achieve via only one lens plane. 
 
\subsection{What modelers can do, in practice}
\begin{figure}
	\centering
	\includegraphics[width=0.5\textwidth]{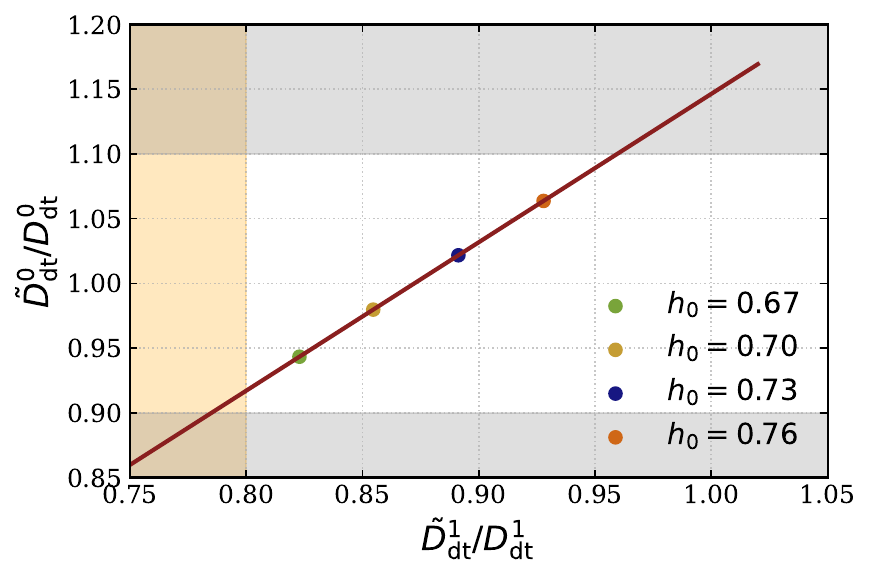}
	\caption{Relation between the convergence factors lying on the time-delay distance $ \tilde{D}_{\rm dt}^n/D_{\rm dt}^n $ and $ h_0 := H_0/(\SI{100}{\kilo\meter\per\second\per\mega\parsec}) $, for the example system described in Sec.~\ref{s:cluster} and Table~\ref{tab:mock}. Bands are regions which would be hypothetically excluded from stellar kinematics measurements with precision $ \delta_\sigma=0.05 $. }
	\label{fig:example}
\end{figure}

From the previous simple example, we can draw the following takeaways:
\begin{itemize}
	\item Multiple plane systems in which unaccounted cores are present on multiple planes are completely consistent, and no fine-tuning is needed between different sheets (unlike what claimed e.g. on Ref.~\cite{Dux:2024vvq}). Fine-tuning claims for the MSD in multiple plane systems are  based on the results of Ref.~\cite{Schneider:2014vka}, which, however, discusses an exact degeneracy, down to non-linear effects on the shears, and without allowing for rescaling of angular diameter distances. We return to this point in Sec.~\ref{s:discussion}.
	\item The role played by stellar kinematics and by the use of shear measurements in breaking the MSD is completely analogous to the single plane lens case.
	\item Measurements of $ \Delta\kappa^{\rm tot}_{Nnm} $ may hint at the presence of unaccounted cores. See Ref.~\cite{Teodori:2022ltt} for a more thorough discussion on this point.
	\item An advantage of multiple plane systems comes from the possibility of finding relations between differential convergence factors across multiple planes, as shown in Figure~\ref{fig:example}. If one has priors on how the cosmological external convergences should behave as a function of redshifts (from e.g. galaxy number count comparisons between observations and cosmological simulations~\cite{Hilbert:2008kb,Johnson:2025lhy}, together with the help of the relation in Eq.~\eqref{eq:unfolding}), together with priors from stellar kinematics, then relations between different (effective) time delay distances can improve constraints on the impact of mass sheets, and ultimately improve the accuracy of the $ H_0 $ inference. In the example of Figure~\ref{fig:example}, one can give priors on the allowed $ \tilde{D}^n_{\rm dt} $ range, hence constraining the $ H_0 $ degeneracy window. 
\end{itemize}
The simplest way for modelers to capture the MSD degeneracy as discussed in this work is to use the MSD-reduced model for both the lens equation and the time delays, Eq.~\eqref{eq:lens_tilde} and Eq.~\eqref{eq:deltaT} respectively, but, let the $ N(N-1)/2 $ $ \tilde{C}_{nm} $ and $ N $ $ \tilde{D}^n_{\rm dt} $ quantities be free parameters. %Then, stellar kinematics can constrain convergence values within $ \lambda_n^{\rm \alpha} $ in Eq.~\eqref{eq:lambda_alpha}, whereas cosmological priors and the unfolding relation Eq.~\eqref{eq:unfolding} can constrain external convergence values. 
We can then distinguish two cases:

\begin{itemize}
	\item \textit{Cosmological parameters are assumed known, with the exception of $ H_0 $ which we want to measure.} This is the easiest path, as $ \Delta\kappa^{\rm tot}_{Nnm} $, Eq.~\eqref{eq:diffkappa}, become observables. After the lens model inference pipeline has converged to a certain posterior for $ \tilde{D}^n_{\rm dt} $, one can use cosmological priors plus stellar kinematics (see Eq.~\eqref{eq:sigma}), together with the relations coming from the inferred $ \Delta\kappa^{\rm tot}_{Nnm} $, to constrain the factors $ \tilde{D}^n_{\rm dt}/D ^n_{\rm dt}$ in a $ H_0 $ degeneracy diagram, in a similar way of what we showed in Fig.~\ref{fig:example}.
	\item \textit{Cosmological parameters are not assumed, and we want to measure them.} This is harder, as now both $ \Delta\kappa^{\rm tot}_{Nnm} $ terms are degenerate with parameters like $ \Omega_{\rm m} $. Still, once the inference has converged to a certain posterior for both $ \tilde{C}_{nm} $ and $ \tilde{D}^n_{\rm dt} $, the same help from stellar kinematics plus cosmological prior of the previous case applies, with the only difference that $ \Delta\kappa^{\rm tot}_{Nnm} $ can here only be constrained and cannot be actively used. 
\end{itemize}
For the reader's convenience, we summarize in Table~\ref{tab:summary} what are the parameters which can be measured or constrained in multiple source systems and under which conditions.

Within this simplified modeling, we are glossing over the fact that finite core effects can reduce the constraining power of stellar kinematics, see Eq.~\eqref{eq:kin} and discussion around. A proper conservative modeling of the situation would introduce a finite-core parametrization of the internal sheets explicitly (on the lines of what done in Ref.~\cite{Shajib:2023uig} for a single lens plane), but what we described above can be a good first approach, to be refined once uncertainties get below few percent.
%; for $ \tilde{C}_{nm} $, which contains effective angular diameter distance ratios, one can use a prior centered around $ C_{nm} $, the same quantity but with the ``bare'' angular diameter distances, assuming one cares about determining $ H_0 $ only.

We remark that what shown here is just an example on how the degeneracy we described in this work would result when applied to a known double source lens system. We leave proper implementation of the degeneracy we discussed here in a realistic mock system for future work.

\begin{table}
\begin{tabular}{ccccc}
\toprule
   & $ (n,m)=(0,0) $ \ & $ (n,m)=(1,0) $ \ & $ (n,m)=(2,0) $ \ & $ (n,m)=(2,1) $ \\
  \midrule
 $ z_n $ 								& 0.184 & 1.885 & 2.38 		& -		\\
$ \kappa_{\mathrm{c} n} $ 				& 0.05 	& 0.03 	& 	-		& -		\\
$ \kappa_n $ 							& -0.013& 0.03	& 0.04		& -		\\
$ \kappa(z_n, z_m) $ 					& 	-	& 0.016 & 0.025  	& 0.015	\\
$ \Gamma_{\mathrm{c}n} $ 				& 0.025 & 0.015	&  -		&	- 	\\
$ \kappa^{\rm tot}_n $					& -0.013 & 0.08  & 0.12 		& -	\\
$ \kappa^{\rm tot}(z_n, z_m) $  		&	-	& 0.01	& 0.054		& 0.015	\\
$ |\tilde\Gamma_{\mathrm{c}}(z_n,0)|$  	&  -	& 0.027	& 0.045		& - 	\\
$ |\tilde\Gamma_{\mathrm{c}}(z_n,z_m)|$  	&  -	& -		& -			& 0.016 \\
\bottomrule
\end{tabular}
\caption{Example of convergence and shear assignment to a setting resembling ``zigzag'' lens system J1721+8842 discussed in Ref.~\cite{Dux:2024vvq}, and corresponding $ \kappa^{\rm tot}(z_n, z_m)  $ and induced $ |\tilde\Gamma_{\mathrm{c}}(z_n,z_m)| := |\tilde\Gamma(z_n,z_m) - \Gamma(z_n,z_m)/(1-\kappa^{\rm tot}(z_n,z_m))| $ corrections. External convergence assignment is done using the external convergence variance estimate method of Ref.~\cite{Teodori:2022ltt} (and making sure they obey the unfolding relation Eq.~\eqref{eq:unfolding}), whereas the $ \kappa_{\mathrm{c} m} $ assignment is arbitrary.}
\label{tab:mock}
\end{table}

\begin{table}
	\setlength{\tabcolsep}{8pt}
	\begin{tabular}{cp{3cm}p{2cm}p{3cm}p{3cm}}
		\toprule
		Parameter  & Name & Observable in  & Condition  & Degeneracy between? \\
		\midrule
		$\displaystyle \frac{D^{\rm eff}_{\rm A}(z_{N},0)  D^{\rm eff}_{\rm A} (z_n, z_m)}{ D^{\rm eff}_{\rm A}(z_{n},0)  D^{\rm eff}_{\rm A}(z_{N}, z_m)}$ & Effective angular diameter distance ratio, Eq.~\eqref{eq:ang_ratio} & Imaging  & None & Cosmological parameters (except $ H_0 $) and convergences \\[6pt]
		
		$\displaystyle \Delta\kappa^{\rm tot}_{Nnm}$  & Corrected differential external convergence, Eq.~\eqref{eq:diffkappa}  & Imaging & Cosmology, with the only exception of $ H_0 $, known & External and internal convergences \\[20pt]
		
		$\displaystyle \tilde{\Gamma}_{n} \ , \ \tilde{\Gamma}_{nm}$ & Reduced shear, Eq.~\eqref{eq:gamma_n}, Eq.~\eqref{eq:gamma_nm} & Imaging & None & Actual LOS shear $ \Gamma_{n}, \Gamma_{nm}$ and convergences plus internal MSD shears  \\[6pt]
		
		$\displaystyle \tilde{D}_{\rm dt}^n$  & Effective time delay distance, Eq.~\eqref{eq:deltaT} & Time delays and imaging & Knowledge of the $ n+1 $-th source time delay & $ H_0 $ and convergences \\[6pt] 
		
		$\displaystyle \frac{\tilde{D}_{\rm dt}^n }{\tilde{D}_{\rm dt}^m } $  & Effective time delay distance ratio, Eq.~\eqref{eq:unfold_rel} & Time delays and imaging & Knowledge of at least one $\tilde{D}_{\rm dt}^n$ & Same as effective angular diameter distance ratio \\
		\bottomrule
	\end{tabular}
	\caption{Summary of which parameters are observable with imaging, time-delays and under which conditions. We also specify on which type of degeneracy they enter.}
	\label{tab:summary}
\end{table}

%\subsection{Clusters?}
%Answer Grillo concerns about mass sheets inside clusters, different $ \lambda $'s? Generically yes, but within a distribution
%
%Central image and MSD? Flattening of profile should enhance the central image, but does this apply to MSD (which should change all magnifications)? Assuming the main profile does not really have a cusp, otherwise the central image is greatly demagnified anyway.

\section{Higher orders in internal mass sheets} \label{s:non_linear}
In Eq.~\eqref{eq:expand}, we considered internal sheet terms to be of the same order of the external sheets. Being this the case, all the discussion of the previous sections is consistent, as higher orders in internal sheets will mix with higher orders in external sheets, which are neglected in the tidal approximation we used (see App.~\ref{s:der}). %This because the lens equation we wrote in Eq.~\eqref{eq:multiplane} uses the tidal approximation, where higher orders in LSS are neglected (see App.~\ref{s:der}).

One can wonder however what happens in a situation where the internal sheet terms are somewhat bigger than external ones. To see this, we can expand, up to second order in $ \kappa_{\mathrm{c}m} $ and $ \Gamma_{\mathrm{c}m} $ terms, the term in Eq.~\eqref{eq:expand}. Focusing at first on the terms proportional to $ \vec{\theta} $, we have
\begin{align}
\begin{aligned}
\text{Eq.~}\eqref{eq:expand} \overbrace{=}^{\vec{\theta} \text{ terms}}&-\sum_{m=0}^{n-1} \qty( C_{nm} \kappa_{\mathrm{c}m} (\kappa_{nm} + \kappa_m) + \sum_{l=0}^{m-1} C_{nm} C_{ml} \kappa_{\mathrm{c} m} \kappa_{\mathrm{c} l} ) \vec{\theta} \\
&-\sum_{m=0}^{n-1} \qty( C_{nm}  (\kappa_{nm} + \kappa_m)\Gamma_{\mathrm{c}m} + \sum_{l=0}^{m-1} C_{nm} C_{ml} (\kappa_{\mathrm{c} m} \Gamma_{\mathrm{c} l} + \kappa_{\mathrm{c} l} \Gamma_{\mathrm{c} m} )) \vec{\theta} \\
&- \sum_{m=0}^{n-1} \qty( C_{nm}  (\Gamma_{nm} + \Gamma_m)\Gamma_{\mathrm{c}m} + \sum_{l=0}^{m-1} C_{nm} C_{ml} (\Gamma_{\mathrm{c} m} \Gamma_{\mathrm{c} l} )) \vec{\theta} \ ,
\end{aligned}
\end{align}
where the first line relates to corrected convergences (that will be reabsorbed in $ \tilde{C}_{nm} $), the second line relates to corrected shears (that will be reabsorbed in $ \tilde{\Gamma}_{nm} $), whereas the third line contains products of shear matrices. In general, traceless symmetric 2 by 2 matrices can be written as 
\begin{equation}
\Gamma_{A} \Gamma_{B} = a \mathbb{I} + b J \ , \ J =\begin{pmatrix}
	0 & 1\\
	-1 & 0
\end{pmatrix}  \ ;
\end{equation}
the part proportional to the identity would again contribute to the convergence part, whereas the antisymmetric part is a genuine higher order effect, which in principle cannot be reabsorbed on the MSD-reduced model of Eq.~\eqref{eq:msd_multi}\footnote{We do however expect that these terms can be reabsorbed in external shears when one considers higher orders in the tidal approximation. We leave such assessment for future work.}. 
%\luca{What is his effect? Can it be reabsorbed in angle model?}. 

The same structure will arise for the second order terms of Eq.~\eqref{eq:expand} proportional to $ \vec{\alpha}_m $, which read
\begin{align}
\begin{aligned}
\text{Eq.~}\eqref{eq:expand} \overbrace{=}^{\vec{\alpha}_m \text{ terms}}&\sum_{m=0}^{n-2} \qty( \sum_{l=m+1}^{n-1} C_{lm} \kappa_{\mathrm{c}l} (\kappa_{nl} + \kappa_{lm}) + 
\sum_{p=m+1}^{n-2} \sum_{l=m+1}^{n-1} \frac{C_{nl}C_{lp}C_{pm}}{C_{nm}} \kappa_{\mathrm{c}l}  \kappa_{\mathrm{c}p} )C_{nm}\alpha_m \\
&+\sum_{m=0}^{n-2} \qty( \sum_{l=m+1}^{n-1} C_{lm} \Gamma_{\mathrm{c}l} (\kappa_{nl} + \kappa_{lm}) + 
\sum_{p=m+1}^{n-2} \sum_{l=m+1}^{n-1} \frac{C_{nl}C_{lp} C_{pm}}{C_{nm}} (\kappa_{\mathrm{c}l}\Gamma_{\mathrm{c}p} + \kappa_{\mathrm{c}p}\Gamma_{\mathrm{c}l}) )C_{nm}\alpha_m \\
&+\sum_{m=0}^{n-2} \qty( \sum_{l=m+1}^{n-1} C_{lm} \Gamma_{\mathrm{c}l} (\Gamma_{nl} + \Gamma_{lm}) + 
\sum_{p=m+1}^{n-2} \sum_{l=m+1}^{n-1} \frac{C_{nl} C_{lp}C_{pm}}{C_{nm}} \Gamma_{\mathrm{c}l} \Gamma_{\mathrm{c}p} )C_{nm}\alpha_m \ .
\end{aligned}
\end{align}
With the exception of the antisymmetric part of the last line, everything can be reabsorbed in the MSD-reduced model via the second-order correction to Eq.~\eqref{eq:kappatot},
\begin{align}
\begin{aligned}
&\kappa^{(2)}_{\rm tot}(z_n, z_m) = -\qty( \sum_{l=m+1}^{n-1} C_{lm} \kappa_{\mathrm{c}l} (\kappa_{nl} + \kappa_{lm}) + 
\sum_{p=m+1}^{n-2} \sum_{l=m+1}^{n-1} \frac{C_{nl}  C_{lp} C_{pm}}{C_{nm}}\kappa_{\mathrm{c}l} \kappa_{\mathrm{c}p}) + K_{nm} \ , \\  
&\kappa^{(2)}_{\rm tot}(z_n, 0) =-\sum_{m=0}^{n-1} \qty( C_{nm} \kappa_{\mathrm{c}m} (\kappa_{nm} + \kappa_m) + \sum_{l=0}^{m-1} C_{nm} C_{ml} \kappa_{\mathrm{c} m} \kappa_{\mathrm{c} l} ) + K_{n} \ ,
\end{aligned}
\end{align}
where with $ K_{nm} $, $ K_n $ we denote the contribution from the product of shear matrices proportional to the identity.
Equivalently, for the the second-order correction to Eq.~\eqref{eq:gammatot}
\begin{align} \label{eq:higher}
\begin{aligned}
&\Gamma^{(2)}_{\rm tot}(z_n, z_m) = -\sum_{m=0}^{n-2} \qty( \sum_{l=m+1}^{n-1} C_{lm} \Gamma_{\mathrm{c}l} (\kappa_{nl} + \kappa_{lm}) + 
\sum_{p=m+1}^{n-2} \sum_{l=m+1}^{n-1} \frac{C_{nl}C_{lp} C_{pm}}{C_{nm}} (\kappa_{\mathrm{c}l}\Gamma_{\mathrm{c}p} + \kappa_{\mathrm{c}p}\Gamma_{\mathrm{c}l}) ) \ , \\
&\Gamma^{(2)}_{\rm tot}(z_n, 0) = -\sum_{m=0}^{n-1} \qty( C_{nm}  (\kappa_{nm} + \kappa_m)\Gamma_{\mathrm{c}m} + \sum_{l=0}^{m-1} C_{nm} C_{ml} (\kappa_{\mathrm{c} m} \Gamma_{\mathrm{c} l} + \kappa_{\mathrm{c} l} \Gamma_{\mathrm{c} m} )) \ .
\end{aligned}
\end{align}
To give a rough understanding on the magnitude of these terms for our example of Sec.~\ref{s:cluster}, we write in Table~\ref{tab:mock_higher} the equivalent of Table~\ref{tab:mock} but for the second order terms only. Notice that second order corrections have opposite sign with respect to the leading order terms, meaning that the second order corrections tend to decrease the magnitude of corrected convergences, not increase it.

We leave for future work a further investigation of the detectability of the antisymmetric term and its interplay with higher order LOS effects, together with accounting for the LOS flexion~\cite{Duboscq:2024asf} in our formalism. We however note that in a setting where $ \Gamma_{\mathrm{c} m} $ terms are negligible, the degeneracy is exact, since one can write Eq.~\eqref{eq:kappatot} including all the $ N-1 $ higher order $ \kappa_{\rm c} $ terms (generalizing Eq.~\eqref{eq:higher} to $ \kappa^{(m)}_{\rm tot} $, $ m=3,\ldots,N-1 $). We moreover note that using the following extended MSD-reduced model
\begin{equation} \label{eq:J}
\vec{\tilde{\beta}}_n = (\mathbb{I} - \tilde\Gamma_n)\vec{\theta} - \sum_{m=0}^{n-1} (\mathbb{I} - \tilde{\Gamma}_{nm}) \tilde{C}_{nm} (\vec{\tilde{\alpha}}_m(\vec{\tilde{\beta}}_m) + \vec{\tilde{J}}_{\mathrm{c} m}) \ , \ \vec{\tilde{J}}_{\mathrm{c} m} = \lambda^\alpha_m \vec{J}_{\mathrm{c} m} \ , \ \vec{J}_{\mathrm{c} m} := \vec\alpha_{\mathrm{c}m} - \kappa_{\mathrm{c}m} \vec\theta \ , 
\end{equation}
where in $ \vec{J}_{\mathrm{c} m} $ we include all non-convergence terms coming from the displacement angle contribution $ \vec\alpha_{\mathrm{c}m} $ of eventual cores, the degeneracy we discussed can be once again made exact, this time without requiring shear terms to be negligible. %Within this extended MSD-reduced model, the $ H_0 $ degeneracy we discussed in the previous sections is untouched, and one can formally generalize it at every order in $ \kappa_{\mathrm{c}} $. Meaning, the degeneracy becomes exact, since it does not rely on any expansion anymore. 
%It is clear then that, without the issues given by the shear matrices, one can write Eq.~\eqref{eq:kappatot} including all the $ N-1 $ higher order $ \kappa_{\rm c} $ terms (generalizing Eq.~\eqref{eq:higher} to $ \kappa^{(m)}_{\rm tot} $, $ m=3,\ldots,N-1 $) and fall back into the degeneracy we discussed.
The only difference is that one treats shear terms (together with other higher order corrections) coming from the core explicitly, rather than adding them to LOS shears. Nevertheless, the $ \vec{J}_{\mathrm{c} m} $ model, if $  \kappa_{\mathrm{c}m} $ is sufficiently high such that the density model inferred from $ \vec{J}_{\mathrm{c} m} $ implies both overdensities and underdensities in the region where images form, can be deemed unlikely to correspond to a physical configuration. Hence, we will not discuss this possibility further. 

\begin{table}
	\begin{tabular}{ccccc}
		\toprule
		& $ (n,m)=(0,0) $ \ & $ (n,m)=(1,0) $ \ & $ (n,m)=(2,0) $ \ & $ (n,m)=(2,1) $ \\
		\midrule
%		$ z_n $ 								& 0.184 & 1.885 & 2.38 		& -		\\
%		$ \kappa^{(2)}_{\mathrm{c} n} $ 				& 0.05 	& 0.03 	& 	-		& -		\\
%		$ \kappa_n $ 							& -0.02	& 0.03	& 0.04		& -		\\
%		$ \kappa(z_n, z_m) $ 					& 	-	& 0.01 	& 0.025  	& 0.015	\\
%		$ \Gamma_{\mathrm{c}n} $ 				& 0.025 & 0.015	&  -		&	- 	\\
		$ \kappa^{(2)}_{\mathrm{tot} }(z_n) $ &	& 0.0005 & -0.003  &  -	\\
		$ \kappa^{(2)}_{\rm tot}(z_n, z_m) $  		&	-	& 0	& -0.0012		& 0	\\
%		$ |\tilde\Gamma_{\mathrm{c}}(z_n,0)|$  	&  -	& 0.027	& 0.045		& - 	\\
%		$ |\tilde\Gamma_{\mathrm{c}}(z_n,z_m)|$  	&  -	& -		& -			& 0.016 \\
		\bottomrule
	\end{tabular}
	\caption{Same as Table~\ref{tab:mock}, but with the second-order contribution only to convergences (we set $ K_n = K_{nm}=0 $). Notice that the second order corrections have opposite sign with respect to the leading order.}
	\label{tab:mock_higher}
\end{table}

\section{Comparison with previous works} \label{s:discussion}
We review here the existing literature on the multiple-plane MSD and compare it with our work.
Ref.~\cite{Schneider:2014vka} (see also comment on Ref.~\cite{McCully:2013fga}), often cited when claiming that MSD requires fine-tuning in multiple plane MSD, discusses a version of the multiple plane MSD which is very similar to the one discussed here, with two critical differences: it does not allow for the rescaling of angular diameter distances (in our language, they set $ \lambda_{nm} =1 $ always) and it is exact on the internal mass sheets. This results in a rather complicated structure of the multiple plane MSD, where internal sheets in the different planes have to obey specific relations, and sometimes be completely absent. We believe that, when discussing MSD in the context of observational impact rather than a purely mathematical degeneracy, the way of Ref.~\cite{Schneider:2014vka} can be too restrictive. 
In reality, we do not have measurements of angular diameter distances dressed with external convergences, and the moment internal sheets have convergence values of the same order of external convergences, internal sheets and external sheets cannot be really separated (only stellar kinematics, or any way to have a prior on the mass distribution content of the lens alone, can discriminate between the two, via finite core effects of internal sheets, see Eq.~\eqref{eq:kin} for an example). 
Only when dealing with settings where $ \Gamma_{\mathrm{c} l}  $ terms are significantly larger than external shears, then one should not discard higher orders in $ \Gamma_{\mathrm{c} l} $ factors, as Ref.~\cite{Schneider:2014vka} correctly does. However, as we discuss in Sec.~\ref{s:non_linear}, if one allows for the explicit presence of cores, convergence factors can still be rescaled away and the degeneracy we discussed here becomes exact.
%We leave for future works the investigation of a MSD where $ \kappa_{\mathrm{c} l} $ higher orders are not neglected like in Ref.~\cite{Schneider:2014vka} but rescaling of angular diameter distances is allowed, like we do here.

Refs.~\cite{Liesenborgs:2008ty,Liesenborgs:2012pu} discuss a particular version of a multiple plane MSD with a density profile which is designed to cause deflection in a source but not in the other, making the MSD exact even when not accounting for the differential convergences. %See Fig. 4 of Ref..~\cite{Liesenborgs:2012pu} 
However, time delays in multiple plane systems completely break this degeneracy, and it has no real impact on the $H_0$ inference bias, since the time-delay surfaces $D_{\rm dt}^n$ do not get rescaled by constants. %As the same authors of Refs.~\cite{Liesenborgs:2008ty,Liesenborgs:2012pu} suggest, 
In any case, we would not call such a degeneracy MSD, since the density profile in Refs.~\cite{Liesenborgs:2008ty,Liesenborgs:2012pu} degeneracy version comes with substructures and ring-like features.

At the same time as this work, Ref.~\cite{Johnson:2026uhh} appeared, which discusses the internal MSD in the context of double source lensing. The conclusions are similar to what shown here, but in this work we give a completely different insight on the problem, together with the generalization to $ N $ sources and the inclusion of shear associated to the internal sheets, absent in Ref.~\cite{Johnson:2026uhh}.

As far as we are aware, there are not other works really discussing the multiple plane MSD. There is however recent effort in characterizing biases from LOS effects~\cite{Wells:2023vgb,Johnson:2024hvl,Duboscq:2024asf,Tang:2025tny,Lin:2025dgn}, and there are many works discussing or attempting to recover cosmological parameters using angular diameter distance ratios~\cite{Collett:2014ola,Sharma:2022xyw,Bowden:2025rph,Grillo:2024rhi}. The latter works do not explicitly take into account biases from mass sheets, primarily because other uncertainties dominate (see however Ref.~\cite{Johnson:2025lhy}, which does take into account external sheets in their inferences). However, the effect of unaccounted mass sheets will have to be taken into account, once measurements become precise at the order of less than $10\%$. This caveat is not discussed here for the first time, as it was addressed in Refs.~\cite{Schneider:2014ifa,Johnson:2025lhy}, but here we point out how the bias on angular diameter distance ratios assumes the form Eq.~\eqref{eq:bias}, which includes internal mass sheets at the same level as external convergence factors.

%\luca{Ref.~\cite{Bradac:2004gc} uses weak lensing distribution inference coming from source populations at different redshifts to break MSD in galaxy clusters. Old work, but check if I can say something meaningful with differential convergences. Anyway, this ref might go out from the discussion.}

%\cite{McCully:2016yfe} (seem not to look for MSD, but read paper anyway), \cite{Fleury:2021tke} (again no real discussion of MSD, but read better).

%\luca{Is everything obvious, since I am considering only first order in everything? Try second order with mathematica, or consider kappac as first order and external stuff as higher order or something like that}
%%%%%%%%%%%%%%%%%%%
\section{Summary}\label{s:sum}
Multiple plane lens systems are an interesting avenue for exploring different regimes of galaxy lensing, and they promise to deliver new insights into the density profiles and structures of lens galaxies. However, as noted in previous works in the literature, multiple plane systems can only marginally help in mitigating the MSD, a primary degeneracy affecting time delay cosmography. 

In this work, we outlined a framework to deal with an approximate multiple plane MSD. The approximate arises from neglecting higher order lens-lens couplings involving the internal mass sheets (while lens-lens couplings are otherwise considered). Our formalism is thus limited to internal mass sheets whose associated shear are $ \mathcal{O}(0.1) $ or smaller (for internal sheets where constant convergence is the only relevant term, the degeneracy is exact). In practice, this is not a significant limitation, as convergence contributions (and hence the parametrically related shear contributions) from internal mass sheets at a level higher than $\mathcal{O}(0.1) $  can already be reasonably constrained through stellar kinematics measurements. In fact, there is an advantage in considering an approximate degeneracy, since restricting the analysis to mathematically exact degeneracies can lead to observationally misleading conclusions, in particular the claim that multiple plane systems necessarily require fine tuning between internal sheets on multiple planes. As we have shown here, this is not the case. 

We have instead presented the appropriate framework on which one should rely when attempting to ameliorate the MSD using multiple sources. This framework critically relies on the determination of effective angular diameter distance ratios, that is angular diameter distances ``dressed'' with mass sheet convergences (both external and internal). The use of effective angular diameter distances makes dealing with the original and the MSD-reduced model almost trivial. 

With hindsight, the only thing that distinguishes internal and external sheets is that the former come from the lens model, and the latter come from background cosmology and LOS effects. But on a mathematical level, they cannot be distinguished, hence if external sheets do contribute to effective angular diameter distances, so must internal sheets.
We remark that the internal mass sheet contributions to these effective angular diameter distance ratios can affect cosmological parameter inference, as also noted in Ref.~\cite{Schneider:2014ifa} (albeit with a different perspective). %Failing to account for possible internal mass sheet contributions to angular diameter distance ratios may bias cosmological parameter inferences. 

We further remark that a key advantage of multiple plane systems lies in the possibility to provide hints about the presence of internal sheets, through differential convergence factors. Differently from Ref.~\cite{Teodori:2022ltt}, we show that explicit lens-lens coupling can increase the number of differential convergences that can be accessed observationally (although more differential convergences are present in the modeling). Nevertheless, the degeneracy between $ H_0 $, the lens model and time delays remains, even in the lucky scenario where time delay measurements on all sources are possible. We therefore emphasize once more that multiple plane systems do not really break the MSD, even when internal sheets in the different planes are not related to each other. Even when higher order effects (beyond plain convergence ones) on cores behaving as internal mass sheets are deemed visible, one can construct a model where the internal cores are explicitly modeled, but the degeneracy in the way we presented it remains otherwise untouched, and becomes exact, as we explain in Sec.~\ref{s:non_linear} (albeit with a particularly artificial core model, which can be difficult to justify in certain regimes). Still, multiple plane systems can help in at least reducing the $ H_0 $ degeneracy window, thanks to cosmological priors on external differential convergences and stellar kinematics measurements on all the lens planes, which can constrain the presence of mass sheets in the system.

%%%%%%%%%%%%%%%%%%%
%
\vspace{0.5cm}
\textit{Acknowledgments.}
We are grateful to P. Fleury, W.J.R. Enzi, C. Grillo for useful discussions which inspired this work. We thank D. Johnson, K. Blum, P. Fleury and M. Millon  for useful comments and remarks on the manuscript. 
%While sending this work for comments prior to submission, we got to know of Ref.~Johnson et al 2026, soon to appear, discussing a closely related topic.
LT acknowledges the support by the European Research Area (ERA) via the UNDARK project (project number 101159929) and the MICINN through the grant ``DarkMaps'' PID2022-142142NB-I00.

\appendix

\section{Multiple plane lens equation derivations} \label{s:der}
We here present some derivations relevant for the multiple plane lens equations. Many of these derivations appear, with varying level of detail, in many works~\cite{Schneider:1992,McCully:2016yfe,Fleury:2020cal,Teodori:2022ltt}, we include them here for completeness.

\subsection{The multiple plane lens equation with LOS effects}
In a perturbed FRWL universe, one can write the lens equation, relating the source angle vector $ \vec{\beta} $ with the observed angle $ \vec\theta $ and the gravitational potential $ \Phi $, as
\begin{equation} \label{eq:lens_full}
	\beta^i(w, \vec{\theta}) = \theta^i -2 \int_{0}^{w} \dd{w'} \frac{f_\kappa (w - w') }{ f_\kappa (w)f_\kappa (w')} \partial_\beta^i \Phi (\vec\beta(w'), w') \ ,
\end{equation}
where $ w  $ denotes comoving distance and $ f_\kappa (w ) $ is the metric distance, which for a flat universe with curvature parameter $ \kappa=0 $, simply reads $ f_\kappa(w) = w $. In particular, the angular diameter distance expression from redshift $ z_m $ to redshift $ z_n $ reads
\begin{equation}\label{eq:D}
D_{\rm A}(z_n,z_m) = \frac{w_n - w_m}{1 + z_n} \ ,
\end{equation}
where the comoving distances $ w_m $ and $ w_n $ correspond to redshift $ z_m $ and $ z_n $ respectively.
In the following we set $ \kappa=0 $ always, hence no confusion with the convergence $ \kappa $ should arise.

With $N$ main deflector planes located at different $w_j$, $j=0,\ldots,N-1$, the gravitational potential reads
\begin{equation}\label{Phi}
	\Phi(\vec\beta(w), w) = \sum_{j=0}^{N-1}\tilde{\Phi}(\vec\beta(w_j), w_j) \delta(w- w_j) + \Phi_{\mathrm{t}} (\vec\beta(w), w) \ ,
\end{equation}
where $ \Phi_{\rm t}(\vec\beta(w), w) $ is the weak (tidal) gravitational potential associated to LSS. We can approximate 
\begin{equation} \label{Phi_tidal}
	\Phi_{\rm t}(\vec\beta(w), w) \simeq \Phi_{\rm t}(0, w) + \beta^i \partial^\beta_i \Phi_{\mathrm{t}} (0, w)  \ .
\end{equation}
Such an approximation goes with the name of tidal approximation, where the LOS column density between different main deflector planes are seen as tidal planes.
Within this approximation, we can write, from Eq.~\eqref{eq:lens_full}, the lens equation integral up to the first deflector sitting at $w_0$ as
\begin{equation} \label{beta_wl}
	\beta^i(w_0) = \theta^i - 2 \int_{0}^{w_0} \dd{w'} \frac{w_0 - w' }{  w_0w'} \partial^\beta_i \Phi_{\rm t} (0, w') - 2 \int_{0}^{w_0} \dd{w'} \frac{ w_0 - w' }{ w_0 w'} \partial^\beta_i \partial^\beta_j \Phi_{\rm t} (0, w') \beta^j(w') \ ;
\end{equation}
the second term on the RHS of the previous is an unobservable overall shift on the displacement angle (independent on $ \vec\theta $), which can be (and will be) reabsorbed in the source coordinates.
Define
\begin{equation}
	A_{ij} (w_2, w_1) := \delta_{ij} - M_{ij}(w_2, w_1) := \delta_{ij} - 2  \int_{w_1}^{w_2} \dd{w'} \frac{ (w_2 - w')  (w' - w_1) }{  (w_2 - w_1)(w')^2} \partial^\beta_i \partial^\beta_j \Phi_{\rm t} (0, w') \ .
\end{equation} 
The $ M_{ij} $ terms are directly connected to weak lensing convergence and shear, which are $ \mathcal{O}(10^{-1} \divisionsymbol 10^{-2}) $. In the tidal approximation, one considers only terms at first order in $ M_{ij} $; in particular, substituting (for $ w' < w_0 $)
\begin{equation}
	\beta^i(w') = \theta^i - 2 \int_{0}^{w'} \dd{w''} \frac{ w' - w'' }{  w' w''} \partial^\beta_i \partial^\beta_j \Phi_{\rm t} (0, w'')\beta^j (w'') 
\end{equation}
in Eq.~\eqref{beta_wl}, we see that at first order in $ M_{ij} $, we should retain only the $ \theta^i $ term of the previous, ending up with
\begin{equation}
	\beta^i(w_0) = A_{ij}(w_0, 0) \theta^j \ .
\end{equation}
For $ w > w_0 $, the situation changes due to the presence of the first strong deflector; from the full $ \Phi $ of Eq.~\eqref{Phi_tidal}, where we don't use the tidal approximation for the strong deflector (just the thin lens approximation), we have, up to $w_1$, 
\begin{align} \label{beta}
	\begin{aligned}
		\beta^i(w_1) =& \theta^i - 2 \int_{0}^{w_1} \dd{w'} \frac{w_1 - w' }{ w_1w'} \partial^\beta_i \partial^\beta_j \Phi_{\rm t} (0, w') \beta^j(w') -2 \frac{w_1 - w_0 }{ w_1w_0} \partial^\beta_i \tilde{\Phi}(\vec\beta(w_0)) \\
		=& A_{ij}(w_1, 0) \theta^j -2 \frac{w_1 - w_0 }{ w_1w_0} \partial^\beta_i \tilde{\Phi}(\vec\beta(w_0)) \\
		&+ 2\int_{w_0}^{w_1} \dd{w'} \frac{w_1 - w'}{ w_1w'} \partial^\beta_i \partial^\beta_j \Phi_{\rm t} (0, w') \qty[ 2 \frac{w' - w_0 }{ w_0w'} \partial^\beta_i \tilde{\Phi}(\vec\beta(w_0)) ] \ ;
	\end{aligned}
\end{align}
on the last step we substituted $ \beta^j(w') $ inside the integral with the term 
\begin{equation}
	\beta^j(w') = \begin{cases}
		&\theta^j  \text{ for } w\leq w_0 \ , \\
		&\displaystyle\theta^j - 2 \frac{ w' - w_0 }{ w_0w'} \partial^\beta_j \tilde{\Phi}(\vec\beta(w_0)) \text{ for } w > w_0 \ , 
	\end{cases}
\end{equation}
so that we always remain at first order in $ M_{ij} $. 
To connect with the conventions adopted in the main text, we define
\begin{equation} \label{eq:alpha_i}
	\vec{\alpha}_n (\vec{\beta}(w_n)) := \frac{w_N - w_n}{w_N w_n}  \partial_j^\beta	\tilde{\Phi}(\vec\beta(w_n)) = \frac{w_N - w_n}{w_N w_n} \hat{\alpha}_j(\vec\beta(w_n)) \ , \ C_{nm} := \frac{(w_n - w_m)w_N}{(w_N-w_m) w_n} \ ;
\end{equation}
with these definitions, the displacement angles $\vec{\alpha}_n (\vec{\beta}(w_n)) $ from the $ n $-th lens plane are normalized with respect to the final source sitting at $w_N$.

Notice that we can rewrite the term in square brackets in Eq.~\eqref{beta} as
\begin{equation}
	2 \frac{w' - w_0 }{ w_0w'} \partial^\beta_i \tilde{\Phi}(\vec\beta(w_0)) = \frac{(w' - w_0) w_1}{w' (w_1 - w_0) } C_{10} \alpha^i_{0}(\vec\beta(w_0)) \ ,
\end{equation} 
and we finally end up with~\cite{Bar-Kana:1995qyu}
\begin{equation} \label{lens_eq weak lensing}
	\beta^i(w_1) = A_{ij} (w_1, 0) \theta^j - A_{ij}(w_1, w_0)C_{10}\alpha^j_{0}\qty(A_{ij}(w_0, 0) \theta^j) \ .
\end{equation}
We can demonstrate the validity of Eq.~\eqref{eq:multiplane} for generic $ n $ by induction. Suppose we know that 
\begin{equation} \label{eq:multi_lens}
	\vec{\beta}(w_n) = A(w_n,0) \vec{\theta} - \sum_{m=0}^{n-1} A(w_n,w_m) C_{nm} \vec{\alpha}_m(\vec{\beta}(w_m)) \ , 
\end{equation}
is valid for $ n $. We want to show the validity of the previous for $ n+1 $. From Eq.~\eqref{eq:lens_full}, using the potential Eq.~\eqref{Phi}, we have
\begin{equation}
	\beta^i(w_{n+1}) = \theta^i - 2 \int_{0}^{w_{n+1}} \dd{w'} \frac{w_{n+1} - w' }{  w_{n+1}w'} \partial^\beta_i \partial^\beta_j  \Phi_{\rm t} (0, w') \beta^j(w') - 2 \sum_{m=0}^{n} C_{(n+1)m} \alpha^i_m(\vec{\beta}_m) \ ;
\end{equation}
we can use the induction step to write, at zeroth order on $ M^{ij}_{nm} := M_{ij}(w_n,w_m) $ tidal factors
\begin{equation}
\vec{\beta}(w) = \vec{\theta} - 2\sum_{m=0}^{k} \frac{w - w_{m} }{  w_{n+1}w_{m}} \grad_{\vec{\beta}}\Phi(\vec{\beta}_m) \ , \ 
k =\begin{cases}
	\text{no sum} & \text{ if } 0 \le w < w_0 \ , \\
	0 & \text{ if } w_0 \le w < w_1 \ , \\
	1 & \text{ if } w_1 \le w < w_2 \ , \\
	\text{etc.}
\end{cases}
\end{equation}
which implies (at first order in tidal quantities)
\begin{align}
\begin{aligned}
-2 \int_{0}^{w_{n+1}} \dd{w'}& \frac{w_{n+1} - w' }{  w_{n+1}w'} \partial^\beta_i \partial^\beta_j  \Phi_{\rm t} (0, w') \beta^j(w')
= -M^{ij}_{n+1} \theta^i + 4 \sum_{m=0}^{n} \int_{w_m}^{w_{m+1}} \dd{w'} \frac{w_{n+1} - w' }{  w_{n+1}w'} \\ 
&\times\sum_{l=0}^{m}  \frac{w' -w_l}{  w_{l}w'}\partial^\beta_i \partial^\beta_j  \Phi_{\rm t} (0, w') \partial_\beta^j \Phi(\vec{\beta}_l) =
 -M^{ij}_{n+1} \theta^j +4 \sum_{l=0}^{n} \int_{w_l}^{w_{n+1}} \dd{w'} \frac{ (w_{n+1} - w')  (w' - w_l) }{  (w_{n+1} - w_l)(w')^2} \\ 
 &\times\partial^\beta_i \partial^\beta_j  \Phi_{\rm t} (0, w') C_{n+1,l} \alpha^j_l(\vec{\beta}_l) 
=  -M^{ij}_{n+1} \theta^i + \sum_{m=0}^{n} M^{ij}_{n+1,m}  C_{n+1,m} \alpha^j_m(\vec{\beta}_m) \ ,
\end{aligned}
\end{align}
which concludes the proof by induction.

\subsection{Multiple planes and stellar kinematics} \label{s:stellar_kin}
The stellar kinematics velocity dispersion on plane $ m $ is expected to obey the spherical Jeans equation, which reads (see e.g. Ref.~\cite{Mamon:2005})\footnote{A more refined non-spherical treatment will not change the scalings we discuss here.}
\begin{equation}
	\frac{1}{\rho^*_m} \dv{(\rho^*_m \sigma_m^2)}{r} + 2\beta^{\rm ani}_m(r) \frac{\sigma_m^2}{r} = -G\frac{M_m(r)}{r^2} \ , 
\end{equation}
where $r$ is the physical distance on the $ m $-th lens plane, $ \sigma_m $ is the radial velocity dispersion and $\beta^{\rm ani}_m$ is the dimensionless parameter describing the velocity anisotropy. $\rho^*_m$ is instead the luminosity density associated to stars, and $ M_m $ is the lens mass enclosed on radius $ r $, all related to the $ m $-th lens plane. The details of $ \rho^*_m $ modeling will not enter in the following. When considering angular variables, one can substitute $ r=D_{\mathrm{A}}(z_m,0) \beta_m = D^{\rm eff}_{\mathrm{A}}(z_m,0) \tilde{\beta}_m $.
%, but we can assume it to be an Hernquist profile,
%\begin{equation}
%	\rho_*(r; a, I_0) = \frac{I_0 a}{2\pi r (r + a)^3}  \ , \ a \simeq 0.5 r_{\rm eff} \ .
%\end{equation}

Following Eq.~\eqref{eq:J}, we can consider as the original model (model 1)
\begin{equation}
\vec{\beta}_n = A_n\vec{\theta} - \sum_{m=0}^{n-1} A_{nm} C_{nm} (\vec{\alpha}_m(\vec{\beta}_m) + \kappa_{\mathrm{c}m}\vec{\beta}_m + \vec{J}_{\mathrm{c} m}(\vec{\beta}_m)) \ ,  %\vec{\tilde{J}}_{\mathrm{c} m} = \lambda^\alpha_m \vec{J}_{\mathrm{c} m} \ , \ \vec{J}_{\mathrm{c} m} := \vec\alpha_{\mathrm{c}m} - \kappa_{\mathrm{c}m} \vec\theta \ , 	\ ,
\end{equation}
where we factor out non-convergence terms of the internal core in $ \vec{J}_{\mathrm{c} m}(\vec{\beta}) = \vec{\alpha}_{\mathrm{c}m }(\vec{\beta}) -\kappa_{\mathrm{c} m}\vec{\beta} $,
whereas the MSD-reduced model (model 2) which is not modeling the internal sheets would read
\begin{equation}
\vec{\tilde{\beta}}_n = \tilde{A}_{n}\vec{\theta} - \sum_{m=0}^{n-1} \tilde{A}_{nm} \tilde{C}_{nm} \vec{\tilde{\alpha}}_m(\vec{\tilde{\beta}}_m)  \ .  
\end{equation}

The stellar kinematics inferred from model 1 will come from the density $ \rho_m + \rho_{\mathrm{c}m}$ and mass $ M_m + M_{\mathrm{c}m} $ profiles inferred from $ \vec{\alpha}_m(\vec{\beta}) + \kappa_{\mathrm{c}m}\vec\beta + \vec{J}_{\mathrm{c} m}(\vec\beta) $ (where $ \rho_{\mathrm{c}m} $ relates to only the internal core $ \kappa_{\mathrm{c}m}\vec\beta + \vec{J}_{\mathrm{c} m}(\vec\beta) $ and same for $ M_{\mathrm{c} m} $), whereas the stellar kinematics inferred from model 2 would come from density $ \tilde{\rho}_m $ and mass $ \tilde{M}_m $ profiles inferred from $ \vec{\tilde{\alpha}}_m(\vec{\tilde{\beta}}_m) $.

To understand the connection between $ \tilde{M}_m $ and $ M_m$, notice that
\begin{equation}
	\vec{\tilde\alpha}_m (\vec{\tilde{\beta}}_m) = \lambda^\alpha_m \underbrace{\frac{1}{\pi} \int \dd[2]{\theta'} \frac{\lambda^{-1}_m \vec{\tilde{\beta}}_m - \vec{\theta}'}{|\lambda^{-1}_m\vec{\tilde{\beta}}_m - \vec{\theta}'  |^2} \kappa_m(\vec\theta')}_{= \vec\alpha_m(\lambda^{-1}_m\vec{\tilde{\beta}}_m)} = 	  \frac{1}{\pi} \int \dd[2]{\theta''} \frac{\vec{\tilde{\beta}}_m - \vec{\theta}''}{|\vec{\tilde{\beta}}_m - \vec{\theta}''  |^2} \underbrace{\frac{\lambda^\alpha_m}{\lambda_m}\kappa_m(\lambda^{-1}_m\vec\theta'')}_{= \tilde\kappa_m(\vec\theta'')} \ ,
\end{equation}	
where we used Eq.~\eqref{eq:lambdas} for connecting $ \alpha_m  $ and $ \tilde\alpha_m $ and we changed variables in the second step, 	$\vec\theta' = \vec\theta''/\lambda_m$. This implies 
\begin{align}
	\begin{aligned}
		\kappa_m( D_m \lambda^{-1}_m\vec\beta) &= \frac{\int \dd{r_z} \rho_m(r_z, D_m \lambda^{-1}_m\vec\beta)}{\Sigma^m_{\rm crit}} = D_m \lambda_m^{-1} \frac{\int \dd{\theta_z} \rho_m(\theta_z,  D_m \lambda^{-1}_m\vec\beta)}{\Sigma^m_{\rm crit}} = \frac{\lambda_m}{ \lambda^\alpha_m} \tilde\kappa_m(D_m\vec\beta) \\
		&=	\frac{\lambda_m}{ \lambda^\alpha_m}  D_m \frac{\int \dd{\theta_z} \tilde\rho_m(\theta_z, D_m\vec\beta)}{\Sigma^m_{\rm crit}} \implies  \rho(\theta_z, D_m \lambda_m^{-1}\vec\beta) = \frac{\lambda^2_m}{ \lambda^\alpha_m} \tilde\rho_m(\theta_z, D_m\vec\beta) \ ,
	\end{aligned}
\end{align}
where $ r_z $ is the proper transverse direction and $ \theta_z $ the respective angular coordinate.
This straightforwardly translates into
\begin{equation}
	M_m(D_m \lambda_m^{-1}\beta) = D^3_m \lambda_m^{-3} \int\dd{\Omega_2}\int_0^\beta \theta'^2\dd{\theta'} \rho_m(\theta_z, D_m \lambda_m^{-1}\vec{\beta}) =  \frac{1}{\lambda_m^\alpha\lambda_m }\tilde{ M}_m(D_m\vec\beta) \ .
\end{equation}
Stellar kinematics is determined by $ \sim GM/r $, as it is evident by looking at the two Jeans equations of model 1 and 2,
\begin{equation}
	(1) \quad \frac{1}{\rho^{*}_m} \dv{(\rho^{*}_m \sigma^2_m(D_m \lambda^{-1}_m\beta))}{\beta} + 2\beta^{\rm ani}_m \frac{\sigma^2_m(D_m \lambda^{-1}_m\beta)}{\beta} = -G\frac{M_m(D_m \lambda^{-1}_m\beta) + M_{\mathrm{c} m}(D_m \lambda^{-1}_m\beta)}{D_m\lambda^{-1}_m \beta^2} \ ,	
\end{equation}
\begin{equation}
	(2) \quad \frac{1}{\rho^{*}_m} \dv{(\rho^{*}_m \tilde\sigma^2_m(D_m \beta))}{\beta} + 2\beta^{\rm ani}_m \frac{\tilde\sigma^2_m(D_m \beta)}{\beta} = -G\frac{\tilde{M}(D_m\beta)}{D_m\beta^2} \ ,	
\end{equation}
where one can easily infer
\begin{equation}
	\sigma^2_m( D_m \lambda^{-1}_m\beta ) = \frac{1}{\lambda_m^\alpha } \tilde\sigma^2_m( D_m \beta ) + \sigma_{\mathrm{c} m}( D_m \beta ) \ ,
\end{equation}
where $ \sigma_{\mathrm{c} m} $ is the part sourced by $ M_{\mathrm{c}m} $. The generalization to the averaged over the line of sight velocity dispersion $ \sigma_{\mathrm{los}, m} $ is straightforward, leading to the result we wrote in Eq.~\eqref{eq:kin}, which could have been guessed just by analogy with the single lens case.

One important comment is in order: we are assuming that the measurement itself of $ \sigma_{\mathrm{los}, m} $ either is not affected by lensing effects of main perturbers, or can be corrected for. In theory, light deflection across a main perturber does not change spectra (which would affect stellar kinematics measurements), apart from uniform (cosmological) redshift which affects all absorption lines equally; in practice, re-weighting across the aperture could introduce biases for example (see e.g. Ref.~\cite{10.1093/mnras/staf112}). We are assuming that such possible biases give uncertainties which are subdominant with respect to the nominal uncertainty without lensing.

%Just for completeness, let's look at the average $\sigma^2_{\rm P}$ expression used in TDCOSMO inferences; we need ($r = D_{\rm L} (1-\kappa^{\rm L})\theta$ for brevity)
%\begin{equation}
%	I_{\rm H}  \sigma^2_{\rm S}(r) := 2\int_{r}^\infty \qty( 1 - \beta_{\rm ani} \frac{r^2}{r'^2} ) \frac{\rho_{*} \sigma^2_r r' \dd{r'}}{\sqrt{r'^2 - r^2}} = \frac{1}{(1-\kappa^{\rm L})^2} \frac{1 - \kappa^{\rm S}}{1 - \kappa^{\rm LS} } \overset{\sim}{(I_{\rm H} \sigma^2_{\rm S}}(D_{\rm L}\theta))
%\end{equation}
%where $I_H$ is the projected Hernquist profile (one can see that $I_{\rm H} (D_{\rm L} (1-\kappa^{\rm L})\vec\theta) = \tilde{I}_{\rm H}(D_{\rm L} \theta)/ (1-\kappa^{\rm L})^2  $  ).
%We end up with the final convolution of the seeing $\mathcal{P}$ over the aperture $\mathcal{A}$
%\begin{equation}
%	\sigma^2_{\rm P} :=  \frac{\int_{\mathcal{A}} I_{\rm H} \sigma^2_{\rm S} (D_{\rm L} (1-\kappa^{\rm L})\vec\theta) * \mathcal{P}  \dd[2]{\theta} }{\int_{\mathcal{A}} I_{\rm H}(D_{\rm L} (1-\kappa^{\rm L})\theta) * \mathcal{P}  \dd[2]{\theta} } = \frac{1 - \kappa^{\rm S}}{1 - \kappa^{\rm LS} } \tilde\sigma^2_{\rm P}	
%\end{equation}
%again, no dependence on $\kappa^{\rm L}$, and decoupling of kinematics coming from imaging only and weak lensing (in particular, the weak lensing correction is just in the normalization $(1 - \kappa^{\rm S})/(1 - \kappa^{\rm LS} ) $).

\subsection{Time delays with LOS and multiple sources} \label{s:tdelay}
The time delay between image solutions of Eq.~\eqref{lens_eq weak lensing} can be computed by exploiting the Fermat principle~\cite{Schneider:1992,Schneider:1997bq,Schneider:2014vka}. First, we can try to recover the term of the time delay which depends on the lensing potential only. Note that, from the perturbed FRWL metric, we can write, for a main deflector at redshift $z_n$
\begin{equation}
	\dv{t}{l} = 1 - 2\Phi \implies \Delta T^{\rm Shapiro}_n = -2\int_{\Delta l_n} \Phi \dd{l} \ ,
\end{equation}
where $l$ is the proper distance at redshift $z_n$ and $\Delta l_n$ is the extension of the deflector. When summing over all the lenses, the time delay as seen by the observer at $w=0$ reads
\begin{equation} \label{eq:shapiro_multi}
	\Delta T^{\rm Shapiro} = -2 \sum_n (1+z_n) \int_{\Delta l_n} \Phi \dd{l} \ ,
\end{equation}
where the factor $(1+z_n)$ takes into account cosmological redshift effects. With the definition Eq.~\eqref{eq:alpha_i} and the corresponding $\grad_{\vec{\beta}} \Psi_n = \vec\alpha_n(\vec\beta)$, we can write
\begin{equation}
	\Psi_n = 2 \frac{w_N - w_n}{w_N w_n} \int_{\Delta w_n} \Phi \dd{w} = 2 \frac{D_{Nn}}{D_N D_n} \int_{\Delta l_n} \Phi \dd{l} \ ,
\end{equation}
where we exploited Eq.~\eqref{eq:D} for the angular diameter distances and $\dd{w_n} = (1+z_n) \dd{l}$. We thus can write Eq.~\eqref{eq:shapiro_multi} as
\begin{equation} \label{time_delay_prefactor}
	\Delta T^{\rm Shapiro} = - \sum_n D_{\rm dt}^n \Psi_n (\vec{\beta}_n) \ , \ D_{\rm dt}^n:= (1+z_n) \frac{D_N D_n}{D_{Nn}} \ . 
\end{equation}
The Fermat principle states that, up to an affine transformation, the lens equation for a single source can be obtained by taking the gradient $\grad_{\vec{\theta}}$ of the light travel time function $\Delta T(\vec{\theta}, \vec{\beta})$ and setting it to zero, granted that the constraint that the end points of the light travel time are fixed is imposed (in other words, that $ \vec\beta  $ is considered as an independent variable). Eq.~\eqref{time_delay_prefactor} can then be used to understand what is the correct prefactor (the affine parameter) entering the light travel time function. For a single lens system, we see that from the function
\begin{align} \label{t_func}
	\begin{aligned}
		\Delta T(\vec{\theta}, \vec{\beta}) &= D_{\rm dt} \Big(\frac{1}{2}\vec\theta^T\left(\mathbb{I}-M(w_{\rm s},0)-M(w_{\rm l},0)+M(w_{\rm s},w_{\rm l}) \right)\vec\theta \\
		&-\vec\beta^T\left(\mathbb{I}-M(w_{\rm l},0)+M(w_{\rm s},w_{\rm l})\right)\vec\theta-\psi((\mathbb{I}-M(w_{\rm l},0))\vec\theta) \Big) \ ,	
	\end{aligned}	
\end{align}
one indeed recovers Eq.~\eqref{lens_eq weak lensing} using $\grad_{\vec{\theta}} \Delta T(\vec{\theta}, \vec{\beta}) = 0$, recalling the definition $\grad_{\vec\theta}{\Psi (\vec\theta)} = \vec{\alpha}(\vec\theta)$. %Notice that Eq.~\eqref{t_func} has the correct prefactor, Eq.~\eqref{time_delay_prefactor}, in front of $\psi ((\mathbb{I} - M^{\rm l})\vec{\theta} )$. Finally, Eq.~\eqref{eq:DtAB} is recovered via $\Delta t_{AB} = t(\vec{\theta}_A, \vec{\beta}) - t(\vec{\theta}_B, \vec{\beta})$.

A trick to write the time delay for multiple lens plane systems is the following (see Refs.~\cite{Schneider:1992,Schneider:2014vka}): consider a light ray which starts from plane $ n+1 $ and gets deflected only on plane $ n $, that is, after being deflected on $ n $, it would go straight to the observer position at $ z=0 $, and hence observed as forming the angle $ \vec\beta_{n} $ on the sky. Generally, this is a kinematically not allowed light ray, as the deflection angle on plane $ n $ might not get you straight to the observer position. Still, the would-be lens equation it would obey to reads (rewritten in a convenient form)
\begin{equation} \label{eq:lens_wouldbe}
\frac{(A^{\rm tot}_{n+1,n})^{-1}}{C_{n+1,n}} (A^{\rm tot}_{n+1} (A^{\rm tot}_n)^{-1} \vec\beta_n - \vec\beta_{n+1}) - \vec\alpha_{n}(\vec\beta_{n})=0 \ ;
\end{equation}
where on $ A^{\rm tot}_{nm}, A^{\rm tot}_n $ we include also the convergence/shear coming from internal core factors, other than the usual LSS ones. Explicitly,
\begin{equation} \label{eq:Atot}
A^{\rm tot}_{nm} := A^{\rm tot}(z_n, z_m) := A(z_n, z_m) - \sum_{l=m+1}^{n-1} \frac{C_{nl} C_{lm}}{C_{nm}}\qty( \kappa_{\mathrm{c} l} \mathbb{I} + \Gamma_{\mathrm{c} l}  ) \ , \ A^{\rm tot}_{n} :=A(z_n, 0) - \sum_{m=0}^{n-1} C_{nm}\qty( \kappa_{\mathrm{c} m} \mathbb{I} + \Gamma_{\mathrm{c} m}  ) \ .
\end{equation}

Following the same reasoning we used for the single lens time delay Eq.~\eqref{t_func}, the time delay corresponding to the lens equation of Eq.~\eqref{eq:lens_wouldbe} is
\begin{equation}
\Delta T_{n+1,n} = D_{\rm dt}^n \qty( \vec{\beta}^\top_n\frac{(A^{\rm tot}_{n+1,n})^{-1}}{C_{n+1,n}} \qty(\frac{1}{2}A^{\rm tot}_{n+1} (A^{\rm tot}_n)^{-1}\vec\beta_n - \vec\beta_{n+1}) - \Psi_n(\vec{\beta}_n) ) + \text{const} \ ,
\end{equation}
where the additive constant is chosen so that the time delay vanishes for the unperturbed segment, corresponding to the situation where $ \alpha_n=0 $ and $ A^{\rm tot}_{n+1} (A^{\rm tot}_n)^{-1}\vec\beta_n = \vec\beta_{n+1} $. Explicitly,
\begin{equation}
\Delta T_{n+1,n} = D_{\rm dt}^n \qty( \frac{\vec{B}^\top_n A^{\rm tot}_{n+1,n}(A^{\rm tot}_{n+1})^{-1} A^{\rm tot}_{n}\vec{B}_n}{2C_{n+1,n}} - \Psi_n(\vec{\beta}_n) ) \ , \ \vec{B}_n :=  (A^{\rm tot}_{n+1,n})^{-1} ( A^{\rm tot}_{n+1}(A^{\rm tot}_n)^{-1} \vec\beta_n - \vec\beta_{n+1}) \ .
\end{equation}

Now, the full $ \Delta T(\{\vec{\beta}_m\}, \beta_N) $ can be seen as 
\begin{equation}\label{eq:tt}
\Delta T(\{\vec{\beta}_m\}, \beta_N) = \Delta T_{N,N-1} + \Delta T(\{\vec{\beta}_m\}, \beta_{N-1}) \ ,
\end{equation}
where $ \Delta T(\{\vec{\beta}_m\}, \beta_{N-1}) $ is the actual time delay from $ \vec{\beta}_{N-1} $ all the way to the observer, passing through all the $ m< N-1 $ planes. The key insight is that $ \Delta T(\{\vec{\beta}_m\}, \beta_{N-1}) $ is the time delay with respect to the reference segment connecting $ \vec\beta_{N-1} $ and observer, while $ \Delta T_{N,N-1} $ is the time delay with respect to the reference segment $ \vec{\beta}_N $- observer (the same reference segment of the full time delay $ \Delta T(\{\vec{\beta}_m\}, \beta_{N}) $) and sum of the two segments $ \beta_N $-$ \beta_{N-1} $ and $ \vec\beta_{N-1} $- observer. This last segment is shared with the reference segment of $ \Delta T(\{\vec{\beta}_m\}, \beta_{N-1}) $, making the equality of Eq.~\eqref{eq:tt} work. One can iteratively do the same reasoning for $\Delta T(\{\vec{\beta}_m\}, \beta_{N-1})  $, arriving at the final result
\begin{equation}
\Delta T(\{\vec{\beta}_m\}, \beta_N) =\sum_{n=0}^{N-1}  D_{\rm dt}^n \qty( \frac{\vec{B}^\top_n A^{\rm tot}_{n+1,n}(A^{\rm tot}_{n+1})^{-1} A^{\rm tot}_{n}\vec{B}_n}{2C_{n+1,n}} - \Psi_n(\vec{\beta}_n) ) \ .
\end{equation}
 
We can verify that indeed, taking gradients of the previous, one can recover the multiple lens equation Eq.~\eqref{eq:multi_lens}. Notice
\begin{align}\label{eq:grad_timedelay}
\begin{aligned}
\grad_{\vec{\beta}_j} \Delta T &= D_{\rm dt}^j \qty(\frac{\vec{B}_j}{C_{j+1,j}} - \vec{\alpha}_j(\vec{\beta}_j)) - \frac{D_{\rm dt}^{j-1}}{C_{j,j-1}} (A^{\rm tot}_{j})^{-1}A^{\rm tot}_{j-1} \vec{B}_{j-1} \\
&=  \frac{ D_{\rm dt}^j}{C_{j+1,j}} \qty(\vec{B}_j - C_{j+1,j}\vec{\alpha}_j(\vec{\beta}_j) - \frac{(w_{j+1} -w_j) w_{j-1}}{w_{j+1} (w_{j} -w_{j-1})} (A^{\rm tot}_{j})^{-1}A^{\rm tot}_{j-1} \vec{B}_{j-1} )  =0 \ ;	
\end{aligned}
\end{align} 
setting the previous to zero, one obtains a lens equation which relates three planes for $ j>0 $. For $ j=0 $, the equation relates only two planes (there are no $ j-1 $ terms) and one obtains immediately the usual single plane lens equation. Generically, from Eq.~\eqref{eq:grad_timedelay}, one has
\begin{align}
\begin{aligned}
\vec{\beta}_{j+1} &= A_{j+1} \vec{\theta} - C_{j+1,j}A^{\rm tot}_{j+1,j} \vec{\alpha}_j 
- \sum_{m=0}^{j-1} \Big[ A^{\rm tot}_{j+1} (A^{\rm tot}_{j})^{-1} A^{\rm tot}_{jm} C_{jm} \\
	+& \frac{(w_{j+1} - w_{j})w_{j-1}}{w_{j+1} (w_j - w_{j-1})} A^{\rm tot}_{j+1,j}(A^{\rm tot}_{j})^{-1}A^{\rm tot}_{j-1}(A^{\rm tot}_{j,j-1})^{-1} (A^{\rm tot}_{jm} C_{jm} - A^{\rm tot}_{j}(A^{\rm tot}_{j-1})^{-1}A^{\rm tot}_{j-1,m} C_{j-1,m} )   \Big] \vec{\alpha}_m \ ;
\end{aligned}
\end{align}
it is very tedious, but still just algebra, to show that the term in square brackets of the previous is equal to $ A^{\rm tot}_{j+1,m} C_{j+1,m}  $, which proves the equivalence between Eq.~\eqref{eq:grad_timedelay} and Eq.~\eqref{eq:multi_lens}.

%The generalization to multiple lens planes, found recursively via $\grad_{\vec\beta_n} \Delta T(\{\vec{\beta}_j\}, \vec{\beta}_{n+1}) = 0 $ to yield the $n+1$-th equation of the multiple lens equation Eq.~\eqref{eq:multi_lens}, reads~\cite{Schneider:2014vka} %\lcol{TODO CHECK AND MAYBE DERIVE}
%\begin{equation}
%	\Delta T(\vec{\theta}, \{\vec{\beta}_i\}) = \sum_{n=0}^{N-1} \frac{D_{\rm dt}^n}{2C_{n+1,n}}\qty( \vec{B}_n^\top A^{-1}_{n+1,n}A_{n+1}A^{-1}_{n} \vec{B}_n - \Psi_n (\vec{\beta}_n)) \ , \ \vec{B}_n := (A^{-1}_{n+1,n}A_{n+1}A^{-1}_{n} \vec{\beta}_n -A^{-1}_{n+1,n} \vec{\beta}_{n+1}) \ ,
%\end{equation}
%where we used the shorthands $A_{nm}:= A(w_n,w_m)$, $A_n := A(w_n,0)$.
%\luca{Check it}

\subsection{Internal mass sheets in the lens equation}
Inserting mass sheets on the various planes inserts $ \kappa_{\mathrm{c} m} $, $ \Gamma_{\mathrm{c}m} $ terms on the $ m $-th planes; if we decide to normalize these $ \kappa_{\mathrm{c} m} $, $ \Gamma_{\mathrm{c}m} $ with the same normalization for the displacement angle of Eq.~\eqref{eq:alpha_i}, then they enter in the way we wrote them on Eq.~\eqref{eq:multiplane} of the main text. To explicitly show it, notice
\begin{equation} \label{eq:kn}
	\kappa_n = \frac{\div\vec\alpha_n}{2} = \frac{w_N -w_n}{w_Nw_n} \int_{\Delta w_n} \dd{w'} \grad^2_\beta ( \Phi(w', \vec\beta) \delta(w' - w_j) )
	= \frac{w_N -w_n}{w_Nw_n} 4\pi G w^2_n a \int_{\Delta x_{\rm pr}} \dd{x_{\rm pr}} \rho(\vec{x}_{\rm pr}) \ ,
\end{equation}
where we used, with $ a $ as the FRWL scale factor, $ \vec{x} $ as the comoving coordinate and $ \vec{x}_{\rm pr} $ as the proper coordinate,
\begin{equation}
	\grad^2_x \Phi = 4\pi G a^2\rho \ , \ \vec{x}_{\rm pr} = a\vec{x} \ , \ \frac{\grad^2_\beta \Phi}{w^2} = \grad^2_x \Phi - \pdv{^2\Phi}{w^2} \ ;
\end{equation}
in the laplacian decomposition, the $ \pdv*{^2\Phi}{w^2} $ term vanishes as a boundary term. Using angular diameter distances, with $ 1/a = 1+z $, we can rewrite Eq.~\eqref{eq:kn} as
\begin{equation}
	\kappa_n = 4\pi G \frac{D_{\rm A}(z_N, z_n) D_{\rm A}(z_n,0) }{D_{\rm A}(z_N,0)} \Sigma (z_n) = \frac{\Sigma (z_n)}{\Sigma_{\rm crit}(z_n)} \ ,
\end{equation}
which matches the factor and conventions of Eq.~\eqref{eq:sigmac}.

\subsection{Cosmological external convergence expression}
The matrices $M(w_2,w_1) $ introduced in this appendix encode external convergence and shear coming from LSS between comoving distance $w_1$ and $w_2$. To make this apparent, we can write them in the following way
\begin{equation}
	\mathbb{I}- M(w_2,w_1) %=:	(1 - \kappa(w_2,w_1)) \Gamma(w_2,w_1) 
	=: 
	\begin{pmatrix}
		1 - \kappa(w_2,w_1) + \gamma_1(w_2,w_1) & \gamma_2(w_2,w_1) \\
		\gamma_2(w_2,w_1) & 1 - \kappa(w_2,w_1) - \gamma_1(w_2,w_1)
	\end{pmatrix} \ ,
\end{equation}
where $\kappa(w_2,w_1)$, $\gamma_i(w_2, w_1)$ are the external convergence and shears between metric comoving distances $w_1$ and $w_2$. Explicitly,
\begin{equation}
	\kappa =  \frac{1}{2} (M_{11} + M_{22}  ) \ , \; \gamma_2  = M_{12}  \ , \;  \gamma_1  = \frac{1}{2} (M_{11} - M_{22}  ) \ .  
\end{equation}
In particular,
\begin{equation}
\kappa(w_2, w_1) = \int^{w_2}_{w_1} \dd{w} \frac{(w_2 -w)(w-w_1)}{(w_2 -w_1)w^2} \laplacian_\beta \Phi \ .
\end{equation} 

We want now to find an expression for $\kappa(w_2,w_1)$ given a (flat $ \Lambda $CDM) cosmology. Use
\begin{equation} \label{eq:Poisson}
 \laplacian_{\vec{x}}{\Phi} = 4\pi G a^2 \rho_{\rm m} \delta = \qty(\frac{3H^2}{2} - 4\pi G \rho_{\Lambda})\delta =  \frac{3H_0^2 \Omega_{\rm{m},0}}{2} a^{-3} \delta \ ,
\end{equation}
where $ \rho_{\rm m} $ is the matter cosmological background density and $ \delta $ is the matter density contrast. The second and third equalities use the first Friedmann equation for a FRWL universe dominated by dark matter and a cosmological constant.
Notice
\begin{equation}
\int^{w_2}_{w_1} \dd{w} \frac{(w_2 -w)(w-w_1)}{w_2 - w_1}\pdv{^2 \Phi}{w^2} = \frac{(w_2 -w)(w-w_1)}{w_2 - w_1}\eval{\pdv{\Phi}{w}}^{w_2}_{w_1} -\frac{1}{w_2 -w_1}  \int \dd{w} (w_2+w_1 - 2w) \pdv{\Phi}{w} \ ; 
\end{equation}
the first term on RHS of the previous is zero while the last term is at most of the same order as $ \Phi \sim 10^{-5}  $ and thus can be neglected in the tidal approximation. Hence, the part of the laplacian related to the $ w $ derivatives does not contribute, and
we can connect the $ \grad^2_\beta $ laplacian to the Poisson equation in Eq.~\eqref{eq:Poisson} and finally write the cosmological convergence in the form of Eq.~\eqref{eq:kappacosmo}.

\subsection{The unfolding relation and internal mass sheets} \label{s:unfolding}
We here prove what claimed in Sec.~\ref{s:time_delay} about the unfolding relation Eq.~\eqref{eq:unfolding} containing no new information as far as $ \kappa_{\mathrm{c}m} $ terms are concerned.

At first order in convergences, and focusing on the $ \kappa_{\mathrm{c}m} $ part only, we can write (we use Eq.~\eqref{eq:deltaT} for $ \tilde{D}_{\rm dt}^i $ and Eq.~\eqref{eq:ang_ratio} for $ \tilde{C}_{ij} $ to express Eq.~\eqref{eq:unfolding} with angular diameter distances, and used the effective angular diameter distance expression in Eq.~\eqref{eq:Deff_def}, where the expression for $ \kappa^{\rm tot} $ can be found in Eq.~\eqref{eq:kappatot})
\begin{align}\label{eq:unfol}
\begin{aligned}
\text{Eq.~}\eqref{eq:unfolding} &\overset{\kappa_{\mathrm{c}m} \text{ part}}{\simeq}  D_N \qty(\sum_{l=0}^{j-1} \underset{\textcolor{orange}{\bullet} \textcolor{blue}{\bullet} \textcolor{purple}{\bullet}  }{\frac{D_{ki} D_{jl}}{D_{ji} D_k D_{Nl}}} \kappa_{\mathrm{c}l}    
+ \sum_{l=i+1}^{k-1} \underset{\textcolor{red}{\bullet}\textcolor{blue}{\bullet}\textcolor{green}{\bullet}}{\frac{D_{j} D_{kl} D_{li}}{D_{ji} D_k D_l D_{Nl}}} \kappa_{\mathrm{c}l} ) 
= D_N\sum_{l=0}^{k-1} \underset{\textcolor{red}{\bullet}\textcolor{blue}{\bullet}\textcolor{purple}{\bullet}\textcolor{green}{\bullet}\textcolor{orange}{\bullet}}{\frac{D_{kl}}{ D_k D_{Nl}}} \kappa_{\mathrm{c}l}\\
&+ D_N\sum_{l=i+1}^{j-1} \underset{\textcolor{blue}{\bullet}}{\frac{D_{jl} D_{li} }{D_{ji} D_l D_{Nl}}} \kappa_{\mathrm{c}l}
+ D_N\frac{1 + z_i}{1+z_j} \qty(\sum_{l=0}^{i-1} \underset{\textcolor{orange}{\bullet}}{\frac{D_{kj} D_{il} }{D_{ji} D_k D_{Nl}}} \kappa_{\mathrm{c}l}
+\sum_{l=j+1}^{k-1} \underset{\textcolor{red}{\bullet}}{\frac{D_{i} D_{kl} D_{lj}}{D_{ji} D_k D_l D_{Nl}} }\kappa_{\mathrm{c}l} ) \ ;
\end{aligned}
\end{align}
where we denoted with $ \textcolor{orange}{\bullet} $ the terms containing $ \kappa_{\mathrm{c}l} $ for $ 0 \le l < i $, with $ \textcolor{blue}{\bullet} $ for $ i<l<j $, with $ \textcolor{red}{\bullet} $ for $ j< l< k $ and finally with $ \textcolor{purple}{\bullet} $ and $ \textcolor{green}{\bullet} $ for $ l=i $ and $ l=j $ respectively. We will show as an example only the $ \textcolor{orange}{\bullet} $ terms; bringing them all to the LHS of Eq.~\eqref{eq:unfol}, we have
\begin{align}
\begin{aligned}
\textcolor{orange}{\bullet} \implies \sum_{l=0}^{i-1}\frac{D_{N}}{D_k D_{Nl}} \qty[\frac{D_{ki} D_{jl}}{D_{ji}} -  D_{kl} -\frac{1 + z_i}{1+z_j} \frac{D_{kj}D_{il}}{D_{ji}}    ]\kappa_{\mathrm{c}l} = 0 \ ;     
\end{aligned}
\end{align}
by explicitly using the expression Eq.~\eqref{eq:D} for the angular diameter distances, one obtains that the term in square brackets is identically zero. Hence, we obtain an uninformative identity.

The same destiny follows for all the other terms, via similar algebra. Hence, all the $ \kappa_{\mathrm{c}m} $ dependence in Eq.~\eqref{eq:unfolding} drops. We remark that this result is completely expected, given the structure in which $ \kappa_{\mathrm{c}l} $ terms appear in $ \kappa^{\rm tot} $, and given that $ \kappa_{\mathrm{c}l} $ terms cannot be influenced by what happens in the other lens planes.

\bibliography{../../bibliography/ref}

\providecommand{\href}[2]{#2}\begingroup\raggedright\begin{thebibliography}{10}

\bibitem{Vegetti:2023mgp}
S.~Vegetti {\em et~al.}, ``{Strong gravitational lensing as a probe of dark
  matter},'' \href{http://arxiv.org/abs/2306.11781}{{\ttfamily arXiv:2306.11781
  [astro-ph.CO]}}.

\bibitem{Refsdal:1964nw}
S.~Refsdal, ``{On the possibility of determining Hubble's parameter and the
  masses of galaxies from the gravitational lens effect},'' {\em Mon. Not. Roy.
  Astron. Soc.} {\bfseries 128} (1964) 307.

\bibitem{Millon:2019slk}
M.~Millon {\em et~al.}, ``{TDCOSMO. I. An exploration of systematic
  uncertainties in the inference of $H_0$ from time-delay cosmography},''
  \href{http://dx.doi.org/10.1051/0004-6361/201937351}{{\em Astron. Astrophys.}
  {\bfseries 639} (2020) A101},
  \href{http://arxiv.org/abs/1912.08027}{{\ttfamily arXiv:1912.08027
  [astro-ph.CO]}}.

\bibitem{Grillo:2024rhi}
C.~Grillo, L.~Pagano, P.~Rosati, and S.~H. Suyu, ``{Cosmography with supernova
  Refsdal through time-delay cluster lensing: Independent measurements of the
  Hubble constant and geometry of the Universe},''
  \href{http://dx.doi.org/10.1051/0004-6361/202449278}{{\em Astron. Astrophys.}
  {\bfseries 684} (2024) L23},
  \href{http://arxiv.org/abs/2401.10980}{{\ttfamily arXiv:2401.10980
  [astro-ph.CO]}}.

\bibitem{Birrer:2018vtm}
S.~Birrer {\em et~al.}, ``{H0LiCOW - IX. Cosmographic analysis of the doubly
  imaged quasar SDSS 1206+4332 and a new measurement of the Hubble constant},''
  \href{http://dx.doi.org/10.1093/mnras/stz200}{{\em Mon. Not. Roy. Astron.
  Soc.} {\bfseries 484} (2019) 4726},
  \href{http://arxiv.org/abs/1809.01274}{{\ttfamily arXiv:1809.01274
  [astro-ph.CO]}}.

\bibitem{Birrer:2020tax}
S.~Birrer {\em et~al.}, ``{TDCOSMO - IV. Hierarchical time-delay cosmography
  joint inference of the Hubble constant and galaxy density profiles},''
  \href{http://dx.doi.org/10.1051/0004-6361/202038861}{{\em Astron. Astrophys.}
  {\bfseries 643} (2020) A165},
  \href{http://arxiv.org/abs/2007.02941}{{\ttfamily arXiv:2007.02941
  [astro-ph.CO]}}.

\bibitem{TDCOSMO:2025dmr}
{\bfseries TDCOSMO} Collaboration, S.~Birrer {\em et~al.},
  \href{http://dx.doi.org/10.1051/0004-6361/202555801}{``{TDCOSMO 2025:
  Cosmological constraints from strong lensing time delays},''} in {\em {3rd
  General Meeting of CosmoVerse}: {Addressing observational tensions in
  cosmology with systematics and fundamental physics}}.
\newblock 6, 2025.
\newblock \href{http://arxiv.org/abs/2506.03023}{{\ttfamily arXiv:2506.03023
  [astro-ph.CO]}}.

\bibitem{Kelly:2023mgv}
P.~L. Kelly {\em et~al.}, ``{Constraints on the Hubble constant from supernova
  Refsdal{\textquoteright}s reappearance},''
  \href{http://dx.doi.org/10.1126/science.abh1322}{{\em Science} {\bfseries
  380} no.~6649, (2023) abh1322},
  \href{http://arxiv.org/abs/2305.06367}{{\ttfamily arXiv:2305.06367
  [astro-ph.CO]}}.

\bibitem{Collett:2014ola}
T.~E. Collett and M.~W. Auger, ``{Cosmological Constraints from the double
  source plane lens SDSSJ0946+1006},''
  \href{http://dx.doi.org/10.1093/mnras/stu1190}{{\em Mon. Not. Roy. Astron.
  Soc.} {\bfseries 443} no.~2, (2014) 969--976},
  \href{http://arxiv.org/abs/1403.5278}{{\ttfamily arXiv:1403.5278
  [astro-ph.CO]}}.

\bibitem{Sharma:2022xyw}
D.~Sharma, T.~E. Collett, and E.~V. Linder, ``{Testing cosmology with double
  source lensing},''
  \href{http://dx.doi.org/10.1088/1475-7516/2023/04/001}{{\em JCAP} {\bfseries
  04} (2023) 001}, \href{http://arxiv.org/abs/2212.00055}{{\ttfamily
  arXiv:2212.00055 [astro-ph.CO]}}.

\bibitem{Bowden:2025rph}
D.~J. Bowden {\em et~al.}, ``{Constraining Cosmology with Double-source-plane
  Strong Gravitational Lenses from the AGEL Survey},''
  \href{http://dx.doi.org/10.3847/1538-4357/ae092e}{{\em Astrophys. J.}
  {\bfseries 993} no.~1, (2025) 124},
  \href{http://arxiv.org/abs/2509.15012}{{\ttfamily arXiv:2509.15012
  [astro-ph.CO]}}.

\bibitem{Falco1985}
E.~E. {Falco}, M.~V. {Gorenstein}, and I.~I. {Shapiro}, ``{On model-dependent
  bounds on H 0 from gravitational images : application to Q 0957+561 A, B.},''
  \href{http://dx.doi.org/10.1086/184422}{{\em apjl} {\bfseries 289} (Feb,
  1985) L1--L4}.

\bibitem{Schneider:2013sxa}
P.~Schneider and D.~Sluse, ``{Mass-sheet degeneracy, power-law models and
  external convergence: Impact on the determination of the Hubble constant from
  gravitational lensing},''
  \href{http://dx.doi.org/10.1051/0004-6361/201321882}{{\em Astron. Astrophys.}
  {\bfseries 559} (2013) A37}, \href{http://arxiv.org/abs/1306.0901}{{\ttfamily
  arXiv:1306.0901 [astro-ph.CO]}}.

\bibitem{Schneider:2014vka}
P.~Schneider, ``{Generalized multi-plane gravitational lensing: time delays,
  recursive lens equation, and the mass-sheet transformation},''
  \href{http://dx.doi.org/10.1051/0004-6361/201424881}{{\em Astron. Astrophys.}
  {\bfseries 624} (2019) A54}, \href{http://arxiv.org/abs/1409.0015}{{\ttfamily
  arXiv:1409.0015 [astro-ph.CO]}}.

\bibitem{Suyu:2009by}
S.~H. Suyu, P.~J. Marshall, M.~W. Auger, S.~Hilbert, R.~D. Blandford, L.~V.~E.
  Koopmans, C.~D. Fassnacht, and T.~Treu, ``{Dissecting the Gravitational Lens
  B1608+656. II. Precision Measurements of the Hubble Constant, Spatial
  Curvature, and the Dark Energy Equation of State},''
  \href{http://dx.doi.org/10.1088/0004-637X/711/1/201}{{\em Astrophys. J.}
  {\bfseries 711} (2010) 201--221},
  \href{http://arxiv.org/abs/0910.2773}{{\ttfamily arXiv:0910.2773
  [astro-ph.CO]}}.

\bibitem{Fleury:2021tke}
P.~Fleury, J.~Larena, and J.-P. Uzan, ``{Line-of-sight effects in strong
  gravitational lensing},''
  \href{http://dx.doi.org/10.1088/1475-7516/2021/08/024}{{\em JCAP} {\bfseries
  08} (2021) 024}, \href{http://arxiv.org/abs/2104.08883}{{\ttfamily
  arXiv:2104.08883 [astro-ph.CO]}}.

\bibitem{Blum:2020mgu}
K.~Blum, E.~Castorina, and M.~Simonovi\'c, ``{Could Quasar Lensing Time Delays
  Hint to a Core Component in Halos, Instead of $H_0$ Tension?},''
  \href{http://dx.doi.org/10.3847/2041-8213/ab8012}{{\em Astrophys. J. Lett.}
  {\bfseries 892} no.~2, (2020) L27},
  \href{http://arxiv.org/abs/2001.07182}{{\ttfamily arXiv:2001.07182
  [astro-ph.CO]}}.

\bibitem{Blum:2021oxj}
K.~Blum and L.~Teodori, ``{Gravitational lensing H0 tension from ultralight
  axion galactic cores},''
  \href{http://dx.doi.org/10.1103/PhysRevD.104.123011}{{\em Phys. Rev. D}
  {\bfseries 104} no.~12, (2021) 123011},
  \href{http://arxiv.org/abs/2105.10873}{{\ttfamily arXiv:2105.10873
  [astro-ph.CO]}}.

\bibitem{Wilson:2016hcs}
M.~L. Wilson, A.~I. Zabludoff, S.~M. Ammons, I.~G. Momcheva, K.~A. Williams,
  and C.~R. Keeton, ``{A Spectroscopic Survey of the Fields of 28 Strong
  Gravitational Lenses: The Group Catalog},''
  \href{http://dx.doi.org/10.3847/1538-4357/833/2/194}{{\em Astrophys. J.}
  {\bfseries 833} (2016) 194},
  \href{http://arxiv.org/abs/1710.09908}{{\ttfamily arXiv:1710.09908
  [astro-ph.GA]}}.

\bibitem{Wilson:2017apg}
M.~L. Wilson, A.~I. Zabludoff, C.~R. Keeton, K.~C. Wong, K.~A. Williams, K.~D.
  French, and I.~G. Momcheva, ``{A Spectroscopic Survey of the Fields of 28
  Strong Gravitational Lenses: Implications for $H_0$},''
  \href{http://dx.doi.org/10.3847/1538-4357/aa9653}{{\em Astrophys. J.}
  {\bfseries 850} no.~1, (2017) 94},
  \href{http://arxiv.org/abs/1710.09900}{{\ttfamily arXiv:1710.09900
  [astro-ph.GA]}}.

\bibitem{Teodori:2023nrz}
L.~Teodori and K.~Blum, ``{Host group degeneracy in gravitational lensing time
  delay determination of H $_{0}$},''
  \href{http://dx.doi.org/10.1088/1475-7516/2023/08/065}{{\em JCAP} {\bfseries
  08} (2023) 065}, \href{http://arxiv.org/abs/2305.19151}{{\ttfamily
  arXiv:2305.19151 [astro-ph.CO]}}.

\bibitem{DiValentino:2021izs}
E.~Di~Valentino, O.~Mena, S.~Pan, L.~Visinelli, W.~Yang, A.~Melchiorri, D.~F.
  Mota, A.~G. Riess, and J.~Silk, ``{In the realm of the Hubble tension-a
  review of solutions},''
  \href{http://dx.doi.org/10.1088/1361-6382/ac086d}{{\em Class. Quant. Grav.}
  {\bfseries 38} no.~15, (2021) 153001},
  \href{http://arxiv.org/abs/2103.01183}{{\ttfamily arXiv:2103.01183
  [astro-ph.CO]}}.

\bibitem{Verde:2023lmm}
L.~Verde, N.~Sch\"oneberg, and H.~Gil-Mar\'\i{}n, ``{A tale of many $H_0$},''
  \href{http://arxiv.org/abs/2311.13305}{{\ttfamily arXiv:2311.13305
  [astro-ph.CO]}}.

\bibitem{DiValentino:2025sru}
E.~Di~Valentino {\em et~al.}, ``{The CosmoVerse White Paper: Addressing
  observational tensions in cosmology with systematics and fundamental
  physics},'' \href{http://arxiv.org/abs/2504.01669}{{\ttfamily
  arXiv:2504.01669 [astro-ph.CO]}}.

\bibitem{Akrami:2018vks}
{\bfseries Planck} Collaboration, N.~Aghanim {\em et~al.}, ``{Planck 2018
  results. I. Overview and the cosmological legacy of Planck},''
  \href{http://dx.doi.org/10.1051/0004-6361/201833880}{{\em Astron. Astrophys.}
  {\bfseries 641} (2020) A1}, \href{http://arxiv.org/abs/1807.06205}{{\ttfamily
  arXiv:1807.06205 [astro-ph.CO]}}.

\bibitem{Abuter:2018drb}
{\bfseries GRAVITY} Collaboration, R.~Abuter {\em et~al.}, ``{Detection of the
  gravitational redshift in the orbit of the star S2 near the Galactic centre
  massive black hole},''
  \href{http://dx.doi.org/10.1051/0004-6361/201833718}{{\em Astron. Astrophys.}
  {\bfseries 615} (2018) L15},
\href{http://arxiv.org/abs/1807.09409}{{\ttfamily arXiv:1807.09409
  [astro-ph.GA]}}.
%%CITATION = ARXIV:1807.09409;%%.

\bibitem{Ivanov:2019pdj}
M.~M. Ivanov, M.~Simonovi\'c, and M.~Zaldarriaga, ``{Cosmological Parameters
  from the BOSS Galaxy Power Spectrum},''
  \href{http://dx.doi.org/10.1088/1475-7516/2020/05/042}{{\em JCAP} {\bfseries
  05} (2020) 042}, \href{http://arxiv.org/abs/1909.05277}{{\ttfamily
  arXiv:1909.05277 [astro-ph.CO]}}.

\bibitem{Troster:2019ean}
T.~Tr\"oster {\em et~al.}, ``{Cosmology from large-scale structure:
  Constraining $\Lambda$CDM with BOSS},''
  \href{http://dx.doi.org/10.1051/0004-6361/201936772}{{\em Astron. Astrophys.}
  {\bfseries 633} (2020) L10},
  \href{http://arxiv.org/abs/1909.11006}{{\ttfamily arXiv:1909.11006
  [astro-ph.CO]}}.

\bibitem{Riess:2021jrx}
A.~G. Riess {\em et~al.}, ``{A Comprehensive Measurement of the Local Value of
  the Hubble Constant with 1 km s$^{-1}$ Mpc$^{-1}$ Uncertainty from the Hubble
  Space Telescope and the SH0ES Team},''
  \href{http://dx.doi.org/10.3847/2041-8213/ac5c5b}{{\em Astrophys. J. Lett.}
  {\bfseries 934} no.~1, (2022) L7},
  \href{http://arxiv.org/abs/2112.04510}{{\ttfamily arXiv:2112.04510
  [astro-ph.CO]}}.

\bibitem{Freedman:2024eph}
W.~L. Freedman, B.~F. Madore, I.~S. Jang, T.~J. Hoyt, A.~J. Lee, and K.~A.
  Owens, ``{Status Report on the Chicago-Carnegie Hubble Program (CCHP):
  Measurement of the Hubble Constant Using the Hubble and James Webb Space
  Telescopes},'' \href{http://arxiv.org/abs/2408.06153}{{\ttfamily
  arXiv:2408.06153 [astro-ph.CO]}}.

\bibitem{Kolatt:1997zh}
T.~S. Kolatt and M.~Bartelmann, ``{Gravitational lensing of type Ia supernovae
  by galaxy clusters},''
  \href{http://dx.doi.org/10.1046/j.1365-8711.1998.01466.x}{{\em Mon. Not. Roy.
  Astron. Soc.} {\bfseries 296} (1998) 763},
  \href{http://arxiv.org/abs/astro-ph/9708120}{{\ttfamily
  arXiv:astro-ph/9708120}}.

\bibitem{Oguri:2002ku}
M.~Oguri and Y.~Kawano, ``{Gravitational lens time delays for distant
  supernovae: break the degeneracy between radial mass profiles and the hubble
  constant},'' \href{http://dx.doi.org/10.1046/j.1365-8711.2003.06290.x}{{\em
  Mon. Not. Roy. Astron. Soc.} {\bfseries 338} (2003) L25--L29},
  \href{http://arxiv.org/abs/astro-ph/0211499}{{\ttfamily
  arXiv:astro-ph/0211499}}.

\bibitem{Grogin:1995gf}
N.~A. Grogin and R.~Narayan, ``{A New model of the gravitational lens 0957+561
  and a limit on the Hubble constant},''
  \href{http://dx.doi.org/10.1086/178171}{{\em Astrophys. J.} {\bfseries 473}
  (1996) 570}, \href{http://arxiv.org/abs/astro-ph/9512156}{{\ttfamily
  arXiv:astro-ph/9512156}}.

\bibitem{Romanowsky_1999}
A.~J. Romanowsky and C.~S. Kochanek, ``Constraints {onH}0from the central
  velocity dispersions of lens galaxies,''
  \href{http://dx.doi.org/10.1086/307077}{{\em The Astrophysical Journal}
  {\bfseries 516} no.~1, (May, 1999) 18--26}.
  \url{https://doi.org/10.1086%2F307077}.

\bibitem{TreuKoopmans02}
T.~{Treu} and L.~V.~E. {Koopmans}, ``{The internal structure of the lens
  PG1115+080: breaking degeneracies in the value of the Hubble constant},''
  \href{http://dx.doi.org/10.1046/j.1365-8711.2002.06107.x}{{\em \mnras}
  {\bfseries 337} no.~2, (Dec, 2002) L6--L10},
  \href{http://arxiv.org/abs/astro-ph/0210002}{{\ttfamily
  arXiv:astro-ph/0210002 [astro-ph]}}.

\bibitem{Foxley-Marrable:2018dzu}
M.~Foxley-Marrable, T.~E. Collett, G.~Vernardos, D.~A. Goldstein, and D.~Bacon,
  ``{The impact of microlensing on the standardization of strongly lensed Type
  Ia supernovae},'' \href{http://dx.doi.org/10.1093/mnras/sty1346}{{\em Mon.
  Not. Roy. Astron. Soc.} {\bfseries 478} no.~4, (2018) 5081--5090},
  \href{http://arxiv.org/abs/1802.07738}{{\ttfamily arXiv:1802.07738
  [astro-ph.CO]}}.

\bibitem{Birrer:2020jyr}
S.~Birrer and T.~Treu, ``{TDCOSMO - V. Strategies for precise and accurate
  measurements of the Hubble constant with strong lensing},''
  \href{http://dx.doi.org/10.1051/0004-6361/202039179}{{\em Astron. Astrophys.}
  {\bfseries 649} (2021) A61},
  \href{http://arxiv.org/abs/2008.06157}{{\ttfamily arXiv:2008.06157
  [astro-ph.CO]}}.

\bibitem{1987ApJ...313..121M}
D.~{Merritt}, ``{The Distribution of Dark Matter in the Coma Cluster},''
  \href{http://dx.doi.org/10.1086/164953}{{\em \apj} {\bfseries 313} (Feb.,
  1987) 121}.

\bibitem{Mamon:2005}
G.~A. Mamon and E.~L. $\L{}$okas, ``{Dark matter in elliptical galaxies — II.
  Estimating the mass within the virial radius},''
  \href{http://dx.doi.org/10.1111/j.1365-2966.2005.09400.x}{{\em Monthly
  Notices of the Royal Astronomical Society} {\bfseries 363} no.~3, (11, 2005)
  705--722}. \url{https://doi.org/10.1111/j.1365-2966.2005.09400.x}.

\bibitem{LSSTScience:2009jmu}
{\bfseries LSST Science, LSST Project} Collaboration, P.~A. Abell {\em et~al.},
  ``{LSST Science Book, Version 2.0},''
  \href{http://arxiv.org/abs/0912.0201}{{\ttfamily arXiv:0912.0201
  [astro-ph.IM]}}.

\bibitem{LSSTDarkEnergyScience:2018jkl}
{\bfseries LSST Dark Energy Science} Collaboration, R.~Mandelbaum {\em et~al.},
  ``{The LSST Dark Energy Science Collaboration (DESC) Science Requirements
  Document},'' \href{http://arxiv.org/abs/1809.01669}{{\ttfamily
  arXiv:1809.01669 [astro-ph.CO]}}.

\bibitem{Euclid:2019clj}
{\bfseries Euclid} Collaboration, A.~Blanchard {\em et~al.}, ``{Euclid
  preparation: VII. Forecast validation for Euclid cosmological probes},''
  \href{http://dx.doi.org/10.1051/0004-6361/202038071}{{\em Astron. Astrophys.}
  {\bfseries 642} (2020) A191},
  \href{http://arxiv.org/abs/1910.09273}{{\ttfamily arXiv:1910.09273
  [astro-ph.CO]}}.

\bibitem{Collett:2015roa}
T.~E. Collett, ``{The population of galaxy-galaxy strong lenses in forthcoming
  optical imaging surveys},''
  \href{http://dx.doi.org/10.1088/0004-637X/811/1/20}{{\em Astrophys. J.}
  {\bfseries 811} no.~1, (2015) 20},
  \href{http://arxiv.org/abs/1507.02657}{{\ttfamily arXiv:1507.02657
  [astro-ph.CO]}}.

\bibitem{Schneider:2014ifa}
P.~Schneider, ``{Can one determine cosmological parameters from multi-plane
  strong lens systems?},''
  \href{http://dx.doi.org/10.1051/0004-6361/201424450}{{\em Astron. Astrophys.}
  {\bfseries 568} (2014) L2}, \href{http://arxiv.org/abs/1406.6152}{{\ttfamily
  arXiv:1406.6152 [astro-ph.CO]}}.

\bibitem{Dux:2024vvq}
F.~Dux {\em et~al.}, ``{J1721+8842: The first Einstein zigzag lens},''
  \href{http://dx.doi.org/10.1051/0004-6361/202452970}{{\em Astron. Astrophys.}
  {\bfseries 694} (2025) A300},
  \href{http://arxiv.org/abs/2411.04177}{{\ttfamily arXiv:2411.04177
  [astro-ph.CO]}}.

\bibitem{Ballard:2023fgi}
D.~J. Ballard, W.~J.~R. Enzi, T.~E. Collett, H.~C. Turner, and R.~J. Smith,
  ``{Gravitational imaging through a triple source plane lens: revisiting the
  \ensuremath{\Lambda}CDM-defying dark subhalo in SDSSJ0946+1006},''
  \href{http://dx.doi.org/10.1093/mnras/stae514}{{\em Mon. Not. Roy. Astron.
  Soc.} {\bfseries 528} no.~4, (2024) 7564--7586},
  \href{http://arxiv.org/abs/2309.04535}{{\ttfamily arXiv:2309.04535
  [astro-ph.CO]}}.

\bibitem{Teodori:2022ltt}
L.~Teodori, K.~Blum, E.~Castorina, M.~Simonovi\'c, and Y.~Soreq, ``{Comments on
  the mass sheet degeneracy in cosmography analyses},''
  \href{http://dx.doi.org/10.1088/1475-7516/2022/07/027}{{\em JCAP} {\bfseries
  07} no.~07, (2022) 027}, \href{http://arxiv.org/abs/2201.05111}{{\ttfamily
  arXiv:2201.05111 [astro-ph.CO]}}.

\bibitem{Schneider:1992}
P.~{Schneider}, J.~{Ehlers}, and E.~E. {Falco},
  \href{http://dx.doi.org/10.1007/978-3-662-03758-4}{{\em {Gravitational
  Lenses}}}.
\newblock Springer, 1992.

\bibitem{McCully:2013fga}
C.~McCully, C.~R. Keeton, K.~C. Wong, and A.~I. Zabludoff, ``{A New Hybrid
  Framework to Efficiently Model Lines of Sight to Gravitational Lenses},''
  \href{http://dx.doi.org/10.1093/mnras/stu1316}{{\em Mon. Not. Roy. Astron.
  Soc.} {\bfseries 443} no.~4, (2014) 3631--3642},
  \href{http://arxiv.org/abs/1401.0197}{{\ttfamily arXiv:1401.0197
  [astro-ph.CO]}}.

\bibitem{Hilbert:2008kb}
S.~Hilbert, J.~Hartlap, S.~D.~M. White, and P.~Schneider, ``{Ray-tracing
  through the Millennium Simulation: Born corrections and lens-lens coupling in
  cosmic shear and galaxy-galaxy lensing},''
  \href{http://dx.doi.org/10.1051/0004-6361/200811054}{{\em Astron. Astrophys.}
  {\bfseries 499} (2009) 31}, \href{http://arxiv.org/abs/0809.5035}{{\ttfamily
  arXiv:0809.5035 [astro-ph]}}.

\bibitem{Johnson:2025lhy}
D.~Johnson, T.~Collett, T.~Li, and P.~Fleury, ``{Line-of-sight effects on
  double source plane lenses},''
  \href{http://dx.doi.org/10.1088/1475-7516/2025/08/067}{{\em JCAP} {\bfseries
  08} (2025) 067}, \href{http://arxiv.org/abs/2501.17153}{{\ttfamily
  arXiv:2501.17153 [astro-ph.CO]}}.

\bibitem{Blum:2024igb}
K.~Blum and L.~Teodori, ``{Hunting for ultralight dark matter with cosmographic
  H0 signal},'' \href{http://dx.doi.org/10.1103/PhysRevD.111.043509}{{\em Phys.
  Rev. D} {\bfseries 111} no.~4, (2025) 043509},
  \href{http://arxiv.org/abs/2409.04134}{{\ttfamily arXiv:2409.04134
  [astro-ph.CO]}}.

\bibitem{Bartelmann:2010fz}
M.~Bartelmann, ``{Gravitational Lensing},''
  \href{http://dx.doi.org/10.1088/0264-9381/27/23/233001}{{\em Class. Quant.
  Grav.} {\bfseries 27} (2010) 233001},
  \href{http://arxiv.org/abs/1010.3829}{{\ttfamily arXiv:1010.3829
  [astro-ph.CO]}}.

\bibitem{McCully:2016yfe}
C.~McCully, C.~R. Keeton, K.~C. Wong, and A.~I. Zabludoff, ``{Quantifying
  Environmental and Line-of-Sight Effects in Models of Strong Gravitational
  Lens Systems},'' \href{http://dx.doi.org/10.3847/1538-4357/836/1/141}{{\em
  Astrophys. J.} {\bfseries 836} no.~1, (2017) 141},
  \href{http://arxiv.org/abs/1601.05417}{{\ttfamily arXiv:1601.05417
  [astro-ph.CO]}}.

\bibitem{Fleury:2020cal}
P.~Fleury, J.~Larena, and J.-P. Uzan, ``{Gravitational lenses in arbitrary
  space-times},'' \href{http://dx.doi.org/10.1088/1361-6382/abea2d}{{\em Class.
  Quant. Grav.} {\bfseries 38} no.~8, (2021) 085002},
  \href{http://arxiv.org/abs/2011.04440}{{\ttfamily arXiv:2011.04440 [gr-qc]}}.

\bibitem{Johnson:2026uhh}
D.~Johnson, P.~Fleury, and M.~Millon, ``{Degeneracies and modelling choices in
  double-plane time-delay cosmography},''
  \href{http://arxiv.org/abs/2602.02697}{{\ttfamily arXiv:2602.02697
  [astro-ph.CO]}}.

\bibitem{Gavazzi:2008aq}
R.~Gavazzi, T.~Treu, L.~V.~E. Koopmans, A.~S. Bolton, L.~A. Moustakas,
  S.~Burles, and P.~J. Marshall, ``{The Sloan Lens ACS Survey. VI: Discovery
  and analysis of a double Einstein ring},''
  \href{http://dx.doi.org/10.1086/529541}{{\em Astrophys. J.} {\bfseries 677}
  (2008) 1046}, \href{http://arxiv.org/abs/0801.1555}{{\ttfamily
  arXiv:0801.1555 [astro-ph]}}.

\bibitem{Collett:2020lii}
T.~E. Collett and R.~J. Smith, ``{A triple rollover: a third multiply imaged
  source at z {\ensuremath{\approx}} 6 behind the Jackpot gravitational
  lens},'' \href{http://dx.doi.org/10.1093/mnras/staa1804}{{\em Mon. Not. Roy.
  Astron. Soc.} {\bfseries 497} no.~2, (2020) 1654--1660},
  \href{http://arxiv.org/abs/2004.00649}{{\ttfamily arXiv:2004.00649
  [astro-ph.CO]}}.

\bibitem{TU:2009cyy}
H.~TU {\em et~al.}, ``{The mass profile of early-type galaxies in overdense
  environments: the case of the double source plane gravitational lens
  SL2SJ02176-0513},'' \href{http://dx.doi.org/10.1051/0004-6361/200911963}{{\em
  Astron. Astrophys.} {\bfseries 501} (2009) 475},
  \href{http://arxiv.org/abs/0902.4804}{{\ttfamily arXiv:0902.4804
  [astro-ph.CO]}}.

\bibitem{Tanaka:2016pdb}
M.~Tanaka {\em et~al.}, ``{A Spectroscopically Confirmed Double Source Plane
  Lens System in the Hyper Suprime-Cam Subaru Strategic Program},''
  \href{http://dx.doi.org/10.3847/2041-8205/826/2/L19}{{\em Astrophys. J.
  Lett.} {\bfseries 826} no.~2, (2016) L19},
  \href{http://arxiv.org/abs/1606.09363}{{\ttfamily arXiv:1606.09363
  [astro-ph.CO]}}.

\bibitem{Schuldt:2019vza}
S.~Schuldt, G.~Chiriv{\`\i}, S.~H. Suyu, A.~Y{\i}ld{\i}r{\i}m, A.~Sonnenfeld,
  A.~Halkola, and G.~F. Lewis, ``{Inner dark matter distribution of the Cosmic
  Horseshoe (J1148+1930) with gravitational lensing and dynamics},''
  \href{http://dx.doi.org/10.1051/0004-6361/201935042}{{\em Astron. Astrophys.}
  {\bfseries 631} (2019) A40},
  \href{http://arxiv.org/abs/1901.02896}{{\ttfamily arXiv:1901.02896}}.

\bibitem{Bolamperti_2023}
A.~Bolamperti, C.~Grillo, R.~Cañameras, S.~H. Suyu, and L.~Christensen,
  ``Reconstructing the extended structure of multiple sources strongly lensed
  by the ultra-massive elliptical galaxy sdss j0100+1818,''
  \href{http://dx.doi.org/10.1051/0004-6361/202244680}{{\em Astronomy \&
  Astrophysics} {\bfseries 671} (Mar., 2023) A60}.
  \url{http://dx.doi.org/10.1051/0004-6361/202244680}.

\bibitem{Peacock:2014}
J.~A. {Peacock} and R.~E. {Smith}, ``{HALOFIT: Nonlinear distribution of
  cosmological mass and galaxies},'' Feb., 2014.

\bibitem{Takahashi:2012em}
R.~Takahashi, M.~Sato, T.~Nishimichi, A.~Taruya, and M.~Oguri, ``{Revising the
  Halofit Model for the Nonlinear Matter Power Spectrum},''
  \href{http://dx.doi.org/10.1088/0004-637X/761/2/152}{{\em Astrophys. J.}
  {\bfseries 761} (2012) 152}, \href{http://arxiv.org/abs/1208.2701}{{\ttfamily
  arXiv:1208.2701 [astro-ph.CO]}}.

\bibitem{Etherington:2023yyh}
A.~Etherington, J.~W. Nightingale, R.~Massey, S.-I. Tam, X.~Cao, A.~Niemiec,
  Q.~He, A.~Robertson, R.~Li, A.~Amvrosiadis, S.~Cole, J.~M. Diego, C.~S.
  Frenk, B.~L. Frye, D.~Harvey, M.~Jauzac, A.~M. Koekemoer, D.~J. Lagattuta,
  M.~Limousin, G.~Mahler, E.~Sirks, and C.~L. Steinhardt, ``Strong
  gravitational lensing's 'external shear' is not shear,'' Nov., 2023.

\bibitem{Khadka:2024bmw}
N.~Khadka, S.~Birrer, A.~Leauthaud, and H.~Nix, ``{Breaking the mass-sheet
  degeneracy in strong lensing mass modelling with weak lensing
  observations},'' \href{http://dx.doi.org/10.1093/mnras/stae1832}{{\em Mon.
  Not. Roy. Astron. Soc.} {\bfseries 533} no.~1, (2024) 795--806},
  \href{http://arxiv.org/abs/2404.01513}{{\ttfamily arXiv:2404.01513
  [astro-ph.CO]}}.

\bibitem{Shajib:2023uig}
A.~J. Shajib {\em et~al.}, ``{TDCOSMO. XII. Improved Hubble constant
  measurement from lensing time delays using spatially resolved stellar
  kinematics of the lens galaxy},''
  \href{http://dx.doi.org/10.1051/0004-6361/202345878}{{\em Astron. Astrophys.}
  {\bfseries 673} (2023) A9}, \href{http://arxiv.org/abs/2301.02656}{{\ttfamily
  arXiv:2301.02656 [astro-ph.CO]}}.

\bibitem{Duboscq:2024asf}
T.~Duboscq, N.~B. Hogg, P.~Fleury, and J.~Larena, ``{Weak lensing of strong
  lensing: beyond the tidal regime},''
  \href{http://arxiv.org/abs/2405.12091}{{\ttfamily arXiv:2405.12091
  [astro-ph.CO]}}.

\bibitem{Liesenborgs:2008ty}
J.~Liesenborgs, S.~De~Rijcke, H.~Dejonghe, and P.~Bekaert, ``{A generalisation
  of the mass-sheet degeneracy producing ring-like artefacts in the lens mass
  distribution},''
  \href{http://dx.doi.org/10.1111/j.1365-2966.2008.13026.x}{{\em Mon. Not. Roy.
  Astron. Soc.} {\bfseries 386} (2008) 307},
  \href{http://arxiv.org/abs/0801.4255}{{\ttfamily arXiv:0801.4255
  [astro-ph]}}.

\bibitem{Liesenborgs:2012pu}
J.~Liesenborgs and S.~De~Rijcke, ``{Lensing degeneracies and mass
  substructure},''
  \href{http://dx.doi.org/10.1111/j.1365-2966.2012.21751.x}{{\em Mon. Not. Roy.
  Astron. Soc.} {\bfseries 425} (2012) 1772},
  \href{http://arxiv.org/abs/1207.4692}{{\ttfamily arXiv:1207.4692
  [astro-ph.CO]}}.

\bibitem{Wells:2023vgb}
P.~Wells, C.~D. Fassnacht, and C.~E. Rusu, ``{TDCOSMO - XIV. Practical
  techniques for estimating external convergence of strong gravitational lens
  systems and applications to the SDSS J0924+0219 system},''
  \href{http://dx.doi.org/10.1051/0004-6361/202346093}{{\em Astron. Astrophys.}
  {\bfseries 676} (2023) A95},
  \href{http://arxiv.org/abs/2302.03176}{{\ttfamily arXiv:2302.03176
  [astro-ph.CO]}}.

\bibitem{Johnson:2024hvl}
D.~Johnson, P.~Fleury, J.~Larena, and L.~Marchetti, ``{Foreground biases in
  strong gravitational lensing},''
  \href{http://dx.doi.org/10.1088/1475-7516/2024/10/055}{{\em JCAP} {\bfseries
  10} (2024) 055}, \href{http://arxiv.org/abs/2405.04194}{{\ttfamily
  arXiv:2405.04194 [astro-ph.CO]}}.

\bibitem{Tang:2025tny}
{\bfseries LSST Dark Energy Science, LSST Strong Gravitational Lensing Science}
  Collaboration, X.~T. Tang, S.~Birrer, A.~J. Shajib, N.~Khadka, and H.~J.
  Best, ``{Analyzing line-of-sight selection biases in galaxy-scale strong
  lensing with external convergence and shear},''
  \href{http://dx.doi.org/10.1088/1475-7516/2025/10/043}{{\em JCAP} {\bfseries
  10} (2025) 043}, \href{http://arxiv.org/abs/2506.04201}{{\ttfamily
  arXiv:2506.04201 [astro-ph.CO]}}.

\bibitem{Lin:2025dgn}
S.~Lin, B.~Hu, C.~Wei, G.~Li, Y.~Shu, X.~Er, and Z.~Fan, ``{Impact of
  Large-Scale Structure along Line-of-Sight on Time-Delay Cosmography},''
  \href{http://arxiv.org/abs/2509.26382}{{\ttfamily arXiv:2509.26382
  [astro-ph.CO]}}.

\bibitem{Bar-Kana:1995qyu}
R.~Bar-Kana, ``{Effect of large scale structure on multiply imaged sources},''
  \href{http://dx.doi.org/10.1086/177666}{{\em Astrophys. J.} {\bfseries 468}
  (1996) 17}, \href{http://arxiv.org/abs/astro-ph/9511056}{{\ttfamily
  arXiv:astro-ph/9511056}}.

\bibitem{10.1093/mnras/staf112}
W.~Xu, D.~Xu, X.~Er, and J.~Ge, ``Biases in galaxy spectral analysis from
  strong lensing differential magnification effect and correction methods,''
  \href{http://dx.doi.org/10.1093/mnras/staf112}{{\em Monthly Notices of the
  Royal Astronomical Society} {\bfseries 537} no.~2, (01, 2025) 1115--1129},
  \href{http://arxiv.org/abs/https://academic.oup.com/mnras/article-pdf/537/2/1115/61522183/staf112.pdf}{{\ttfamily
  https://academic.oup.com/mnras/article-pdf/537/2/1115/61522183/staf112.pdf}}.
  \url{https://doi.org/10.1093/mnras/staf112}.

\bibitem{Schneider:1997bq}
P.~Schneider, ``{The Cosmological lens equation and the equivalent single plane
  gravitational lens},'' \href{http://dx.doi.org/10.1093/mnras/292.3.673}{{\em
  Mon. Not. Roy. Astron. Soc.} {\bfseries 292} (1997) 673},
  \href{http://arxiv.org/abs/astro-ph/9706185}{{\ttfamily
  arXiv:astro-ph/9706185}}.

\end{thebibliography}\endgroup
\bibliographystyle{utphys}

\end{document}